%% file: main.tex
\documentclass[11pt,a4paper]{article}

\usepackage[T1]{fontenc}
\usepackage[utf8]{inputenc}
\usepackage{lmodern}
\usepackage[margin=2.5cm]{geometry}
\usepackage{setspace}
\usepackage{amsmath,amssymb}
\usepackage{booktabs}
\usepackage{graphicx}
\usepackage{natbib}
\usepackage{hyperref}
\hypersetup{colorlinks=true, linkcolor=black, citecolor=black, urlcolor=blue}
\usepackage{rotating}
\usepackage{caption}
\usepackage{subcaption}
\usepackage{multirow}
\usepackage{array}
\usepackage{pdflscape}
\usepackage{threeparttable}
\usepackage{xcolor}
\usepackage{appendix}
\usepackage{alphalph}  
\usepackage{longtable}
\usepackage{makecell}
\usepackage{etoolbox}
\usepackage{placeins}
\usepackage{pifont}

\AtBeginEnvironment{table}{\singlespacing\centering\footnotesize}
\AtBeginEnvironment{sidewaystable}{\singlespacing\centering\scriptsize}

\newcommand{\TableBarrier}{\FloatBarrier\vspace{0.25em}}

\newcommand{\cmark}{\ding{51}}

\begin{document}

\begin{titlepage}
\begin{center}
\vspace*{2cm}
{\LARGE\bfseries
  Evaluating Structured Strategy Backtests:\\[0.3em]
  Peer Benchmarks, Regime Timing,\\[0.3em]
  and Live Performance}\\[3em]

{\large Chang Liu$^{\,\text{a},\text{b}}$}\\[0.5em]
{\normalsize
$^{\text{a}}$\,Department of Information Engineering and Computer Science,
University of Trento\\[0.2em]
$^{\text{b}}$\,Resonanz Capital GmbH, Frankfurt am Main, Germany\\[0.5em]
Correspondence: chang.liu@resonanzcapital.com}\\[1.2em]

{\small\itshape
Preprint. Under review at the \textit{Journal of Asset Management}.}\\[0.5em]

\vfill

{\footnotesize
\textbf{Disclosure.} The author is employed by Resonanz Capital GmbH, an
institutional asset manager that provided access to the proprietary
strategy-level dataset analysed in this paper.
Resonanz Capital has no financial interest in the specific strategies
studied and did not participate in the design of the empirical tests,
the analysis, or the drafting of the manuscript.
The views expressed here are those of the author and do not necessarily
represent the views of Resonanz Capital, the University of Trento, or
any of their affiliates.
The author declares no other conflicts of interest.
}

\end{center}
\end{titlepage}

\newpage
\begin{abstract}
\noindent
Institutional allocators often evaluate structured strategies on the
basis of marketed backtests---hypothetical track records constructed by
applying a strategy's rules to historical data prior to any live
trading, also referred to as pro-forma performance.
It is unclear how much of that signal survives once the strategy is
actually traded.
Using 1,726 commercially distributed structured strategies from ten
global institutions, this paper shows that raw pro-forma performance
has only limited portability into the live period and weakens sharply
once live outcomes are measured relative to peer and external
benchmarks.
The evidence indicates that marketed backtests predominantly reflect
the common factor regime present before launch rather than
strategy-specific skill.
Strategies launched after unusually strong bucket-factor conditions
experience materially worse subsequent deterioration.
For allocators, the implication is practical: backtests should be
judged relative to appropriate peer benchmarks, and the discount
applied to them should increase when launch occurs after an extreme
factor run.



\end{abstract}

\section{Introduction}
\label{sec:introduction}

Consider an institutional investor evaluating a pitch for a new
structured strategy.
The marketing deck presents five years of daily returns: an
annualised Sharpe ratio exceeding 1.0, shallow drawdowns, and consistent
outperformance across market regimes.
What the deck does not emphasise, however, is that the entire
track record is pro-forma---constructed by applying the strategy's rules
to historical data before a single dollar was traded.
The allocator's dilemma is immediate: how much of that polished record
will survive contact with live markets, and does the apparent signal
reflect genuine skill or merely a favourable environment for the
underlying risk factor?

As used in this paper, ``structured investment strategies'' refers to
commercially distributed, rules-based institutional strategies
marketed by banks and asset managers on the basis of pro-forma (backtested)
track records.
They are distinct from retail structured notes (capital-protected or
barrier products sold to individuals) and from published academic
factor portfolios.
The typical end user is an institutional allocator---a pension fund,
sovereign wealth fund, or family office---evaluating launch-ready
strategies for portfolio inclusion or as a risk-management overlay.

Despite the practical importance of evaluating these products,
the question of pro-forma reliability has received
surprisingly little empirical attention.
The backtest-overfitting literature demonstrates that academic factors
lose predictive power after publication
\citep{Harvey2016, McLean2016, Bailey2014}, and the fund-performance
literature documents that hedge fund alpha declines after
registration \citep{Jagannathan2010, Baquero2005}.
Neither body of work, however, addresses the specific product that
allocators purchase in practice: structured strategies distributed by banks
and asset managers on the strength of a backtested pitch book.
These products occupy a gap between academic factor anomalies
and actively managed funds---rules-based like the former,
yet commercially marketed like the latter.
The underlying return sources---value, momentum, carry, and related
premia---are well documented in the cross-section of expected returns,
but neither the commercial packaging nor the launch-timing decision
has received comparable study.
No existing study, however, examines whether the pro-forma track records of
these commercially distributed strategies retain predictive content once the
common factor environment is accounted for.

This paper examines the question using a proprietary dataset of 1,726
commercially distributed structured strategies from ten global
institutions, spanning the period 2009--2025.
The strategies cover equities, rates, foreign exchange, credit, and
commodities.
Each strategy is assigned to one of seven benchmarking
buckets---Carry, Hedging, Momentum, Multi Premia, Factor, Value, or
Liquidity---that group strategies sharing a common return-generating
mechanism.
Broader objective groups (Return-Seeking,
Hedging/Defensive, Carry/Short-Convexity, Multi-Premia/Diversifying)
capture each strategy's intended portfolio role.

The raw data reveal substantial decay.
Volatility-adjusted returns decline by 2.1--3.1 percentage points from
the pro-forma to the live period, and a moving-block bootstrap confirms
that the magnitude exceeds what mechanical mean reversion alone would
produce.\footnote{Institution-level wild-cluster bootstrap checks
(Appendix~\ref{app:wild_bootstrap}) confirm these results; the ten
anonymised institutions are the natural cluster units for inference
about commercial dissemination of structured strategies, and the
Webb 6-point wild-cluster bootstrap of \citet{Cameron2015} is used
to accommodate the small number of clusters.}
The more consequential question, however, concerns the source of this
decay.
To investigate, we construct a leave-one-out (LOO) bucket-average
benchmark for each strategy---defined as the equal-weighted daily return
of all strategies in the same bucket, excluding the strategy itself
(Section~\ref{sec:methods_bench}).
Once live performance is measured relative to this peer benchmark,
the predictive content of the marketed backtest is substantially
attenuated: the residual association between pro-forma and live
returns, after conditioning on the bucket factor, is economically
negligible relative to the unconditional estimate, indicating that
the marketed backtest predominantly captures a favourable pre-launch
factor regime rather than strategy-specific skill.
An independent Bloomberg total-return benchmark encompassing all
1{,}726 strategies corroborates this finding.

Two structural channels account for this absorption.
First, regime timing: strategies launched when the bucket factor is
substantially above its long-run mean embed an unusually favourable
environment that is unlikely to persist.
The spread in post-launch decay between the coldest and hottest regime
quintiles exceeds four percentage points.
Second, launch-cohort density---a proxy for crowding.
Log launch density does not predict average decay directly at either
horizon, but it moderates the predictive content of the pro-forma
signal at the six-month horizon, with no comparable effect at twelve
months.
We interpret the horizon asymmetry as consistent with short-lived
capacity erosion that dissipates as crowded positions are unwound;
the evidence is therefore consistent with a limited, horizon-dependent
crowding channel rather than a level effect on decay.

This paper makes three contributions.
First, it quantifies the gap between pro-forma and live performance on
a uniquely large commercial sample (1,726 strategies, ten institutions,
2009--2025), closing a gap left by the backtest-overfitting and
fund-persistence literatures.
Second, it introduces a leave-one-out (LOO) bucket-average peer
benchmark that absorbs the common-factor component of pro-forma
performance and isolates strategy-specific skill, showing that the
residual information content of a marketed backtest is economically
negligible once this benchmark is applied.
Third, it documents two structural channels---regime timing at launch
and a horizon-dependent launch-density effect---that jointly explain
the residual decay, and translates these results into an operational
due-diligence rule: the haircut applied to a marketed backtest should
rise with the extremity of the factor regime at launch.
Together, these findings give institutional allocators an
evidence-based framework that dominates raw-backtest screening.

The remainder of the paper is organised as follows.
Section~\ref{sec:literature} reviews the related literature.
Section~\ref{sec:data} describes the data and research design.
Section~\ref{sec:results} presents the empirical results, including
benchmark-conditioning tests and the regime-timing and crowding
channels.
Section~\ref{sec:implications} discusses the implications for
allocators, and Section~\ref{sec:conclusion} concludes.

\section{Related Literature}
\label{sec:literature}

That backtests overstate live performance is among the oldest
lessons in quantitative finance, yet the problem has proven remarkably
persistent.
\citet{Harvey2016} demonstrated that the volume of
published factor research implies a multiple-testing burden so severe
that conventional significance thresholds cannot reliably
distinguish genuine predictors from false discoveries.
\citet{Bailey2014} formalised the mechanism by which even a single
optimised backtest inflates out-of-sample statistics, and
\citet{McLean2016} provided large-scale empirical confirmation:
published cross-sectional predictors decay significantly after
publication, consistent with a combination of data mining and investor
learning.
\citet{Jensen2023} extended this result to 153 factors across
93 countries, finding that the majority replicate out of sample,
although the strongest factors exhibit post-publication attenuation
consistent with rational learning about the true premium rather than
with pure data mining.

These findings, however, leave open a deeper question: what happens to
the signal that remains?
\citet{Penasse2022} offers a structural answer.
His model of alpha decay shows that crowding and information diffusion
generate progressively decelerating performance erosion even for
genuine sources of alpha, and that the erosion is more pronounced when
the pre-discovery signal is stronger.
\citet{BerkGreen2004} arrive at a complementary prediction from the
capacity side: as additional capital flows toward a given source of
alpha, returns converge to the breakeven level, implying that the
density of competing strategies should predict inferior outcomes.
\citet{Pedersen2009} formalises how near-simultaneous entry by multiple
providers into the same factor space amplifies this capacity erosion.
\citet{MitchellPedersen2007} provide a microfoundation for this
mechanism by documenting that capital is slow to move toward
attractive trades even when the opportunity is observable, so that
crowding can persist long enough to generate measurable decay in the
affected cohort.
\citet{ContWagalath2013} formalise how correlated deleveraging among
investors with overlapping positions amplifies these effects and
generates endogenous return correlations on a launch-cohort
time-scale.
\citet{Arnott2016} apply this logic directly to factor strategies,
confirming that strategies launched after periods of historically
strong premia tend to underperform.
Collectively, these papers predict not only that backtested performance
will decay, but that the rate of decay should depend on two observable
conditions at launch: the intensity of the prevailing factor
environment---measured by the deviation of the underlying risk premium
from its long-run average---and the degree of crowding, captured by
the number of competing strategies entering the same risk-premium
space in a given period.
Both hypotheses are tested in
Sections~\ref{sec:regime_extremity} and~\ref{sec:launch_density}.

A related but distinct mechanism, operating on the selection margin
rather than on post-launch dynamics, further widens the wedge between
backtested and live performance.
Product providers tend to launch strategies following unusually
favourable factor conditions
\citep[the ``catering'' channel of][]{Celerier2017},
so that the backtest period is systematically unrepresentative of the
long-run return distribution.
Any decay measured on the launched sample therefore represents a
conservative lower bound, because strategies proposed but never
launched---those that failed the institution's internal screen---are
unobserved \citep[cf.\ the mutual-fund incubation analogy
of][]{Evans2010}.

A parallel literature examines whether performance persists in managed
funds.
The evidence accumulated over nearly three decades is predominantly
negative.
\citet{Carhart1997} found little mutual-fund outperformance beyond
momentum; \citet{Baquero2005} documented short-horizon hedge-fund
persistence that evaporates within a year; and \citet{Jagannathan2010}
showed that hedge-fund performance persistence is concentrated among
superior funds and largely disappears once backfill bias is
corrected---suggesting that pre-registration track records
systematically overstate subsequent alpha.
\citet{Greenwood2014} added a behavioural dimension, showing that
extrapolative expectations generate systematic over-optimism in return
forecasts, offering one explanation for why pro-forma bias may survive
investor sophistication.
\citet{Kosowski2006} and \citet{FamaFrench2010} use bootstrap
procedures to separate skill from luck in the cross-section of mutual
fund returns and find that, once sampling variability is accounted
for, the population of active managers exhibits little evidence of
persistent alpha---a cautionary benchmark for inferences about
backtested commercial products whose marketed track records have not
been subjected to comparable scrutiny.
More recent evidence indicates that persistence can emerge in specific
settings, particularly in small portfolios drawn from the extreme
positive tail of past fund performance, underscoring the sensitivity
of persistence results to portfolio construction and evaluation design
\citep{Cuthbertson2023}.

The structured strategies examined in this paper harvest the same
return sources as academic alternative risk premia (ARP) strategies
\citep{Asness2013, Ilmanen2012}.
Alternative strategies expose investors to a wide range of risk
factors, so that measured performance is inseparable from the
framework used to evaluate it \citep{Amenc2003}---an observation that
motivates the benchmark-design question central to the ARP literature.
\citet{Roncalli2017} surveys the ARP landscape and documents
widespread performance disappointment following the 2018--2020 period,
attributing much of the shortfall to factor crowding and regime
dependence.
A central finding of that literature is that ARP benchmark returns
are provider-dependent: indices from different institutions purporting
to capture the same premium can differ by several hundred basis points
per annum (p.a.), rendering a single market index an inadequate benchmark
for heterogeneous strategy samples \citep{Hamdan2016}.
\citet{Blin2021} reinforce this point, demonstrating that macro-regime
conditions explain a substantial share of cross-sectional ARP return
variation.
If the pro-forma window over-weights a favourable factor cycle,
measured decay should attenuate once common-factor exposure is
removed.

Benchmarking heterogeneous strategies is itself a long-standing
problem.
\citet{DGTW1997} introduced characteristic-based benchmarks to
circumvent the misspecification inherent in parametric factor models,
and \citet{Hunter2014} extended the approach to the fund level with
the Active Peer Benchmark (APB)---the equal-weighted average excess
return of all funds in the same style category---demonstrating that it
substantially reduces residual cross-correlation within peer groups.
For hedge funds, \citet{FungHsieh2004} proposed a seven-factor model
capturing option-like payoff profiles, and \citet{Lhabitant2001}
argued more broadly that alternative-strategy evaluation requires risk
frameworks better tailored than a single market index.
\citet{Fieberg2019} compared risk-model and characteristic-model
approaches from an investor perspective, finding that
characteristic-model estimates are the most informative---a result
that strengthens the case for peer-based rather than index-based
benchmarking.
\citet{Cremers2012} demonstrate that many widely used benchmark
indices themselves exhibit non-zero alpha against standard factor
models, implying that the apparent performance of an evaluated
portfolio depends as much on the benchmark's own factor exposures
as on the portfolio's.
\citet{Mateus2019} reach a similar conclusion in the mutual-fund
setting: when benchmarks are better matched to fund objectives,
measured alpha declines and fund rankings shift materially,
demonstrating that benchmark mismatch can mislead investors regarding
relative performance.
The LOO bucket-average benchmark employed here adapts the
\citet{Hunter2014} APB to the structured-strategy setting: the
leave-one-out exclusion eliminates the mechanical correlation between
a strategy's own return and its benchmark.
Relatedly, \citet{Wagner2002} formalises portfolio selection in a
benchmark-relative setting consistent with the allocator perspective
adopted here, in which candidate strategies are evaluated against
alternative uses of risk budget rather than in isolation.

The study closest in spirit is \citet{Falck2022}, who decompose
performance decay across published systematic strategies and find that
post-publication returns decline by approximately 43\%, with overfitting
accounting for the bulk of the reduction.
The present setting differs materially from prior work in three
respects.
First, the unit of analysis is commercially marketed institutional
products rather than published academic anomalies or managed funds, so
the relevant selection mechanism is the institutional launch decision
rather than academic peer review.
Second, the evaluation is peer-relative by construction: a within-bucket
LOO benchmark absorbs the shared factor return, allowing a direct test
of whether the pro-forma signal retains predictive content after this
absorption---a test that neither the post-publication decay literature
nor the fund-persistence literature performs.
Third, the analysis isolates two launch-timing channels---regime
extremity and launch-density crowding---that account for the residual
decay.

In summary, the existing literature establishes that backtests
frequently decay, that persistence results depend on the evaluation
framework, and that benchmark choice can materially alter inferences
about performance.
What remains untested is whether the marketed backtests of
institutionally distributed structured strategies retain useful
information once the common factor environment is accounted for.
That is the question this paper addresses.

\section{Data and Design}
\label{sec:data}

\subsection{Sample}

The sample comprises 1,726 structured investment strategies distributed
by ten global financial institutions (anonymised A--J) to institutional
clients over the period 2009--2025.
As defined in Section~\ref{sec:introduction}, these are commercially
distributed, rules-based strategies marketed on the basis of pro-forma
(backtested) track records to institutional allocators for allocation
or overlay purposes; the sample excludes retail structured notes and
published academic factor portfolios.
The strategies span six asset classes: equities (39\%), commodities
(21\%), rates (14\%), foreign exchange (11\%), credit (8\%), and
multi-asset (7\%) (see Table~\ref{tab:sample_summary}).

For each strategy, the launch date---defined as the date the strategy
first became available for live investment---divides the track record
into a pro-forma period (mean length 13.7 years) and a live period
(mean length 5.6 years).

All 1,726 strategies satisfy the baseline inclusion criteria: at least
one year of pro-forma history and at least six months of live data
with a valid volatility-adjusted return.
No strategy is excluded from the analysis on the basis of poor
post-launch performance, so within-sample attrition does not
contaminate the estimates.
Pre-launch selection, by contrast, remains material: strategies
proposed but never launched by the sponsoring institutions are
unobserved, so the sample is positively selected on the institutional
go/no-go decision.
The decay estimates reported below should therefore be read as a lower
bound on the gap that would prevail in an unselected universe of
candidate strategies (see Section~\ref{sec:conclusion} for the
implications of this positive selection).
For twelve-month regressions, the sample narrows to 1,694 strategies
with at least twelve months of live data; all other analyses use the
full sample of 1,726.

Each strategy is also assigned to a broader portfolio role---
Return-Seeking ($N=457$, 26\%), Carry/Short-Convexity ($N=808$, 47\%),
Hedging/Defensive ($N=300$, 17\%), or Multi-Premia/Diversifying
($N=161$, 9\%).
This role classification is used in Section~\ref{sec:implications}
only to qualify the interpretation of the Hedging/Defensive group,
whose intended payoff profile is not captured by an unconditional
return metric.
The bucket classification, not the role classification, determines
the appropriate peer benchmark and is the operative grouping
throughout the empirical analysis.
Table~\ref{tab:sample_summary} summarises the sample.
Panel~A reports the composition by asset class, objective group, and
bucket; the full institution-level and launch-year breakdown is
reported in Appendix~\ref{app:sample_detail}.
Panel~B reports the distribution of track-record lengths---the
pro-forma period (backtest inception to launch) and the live period
(launch to the end of the observation window)---in calendar years.

All returns in the dataset are reported gross of transaction costs.
Because transaction costs and fees reduce live performance but do not
affect the marketed backtest, the estimated decay should be interpreted
as a lower bound on investors' net experience.

  \begingroup
  \singlespacing
  \scriptsize
  \setlength{\tabcolsep}{3pt}%
  \renewcommand{\arraystretch}{0.90}%
  \input{table1_summary_main}  \endgroup

\TableBarrier

\subsection{Performance measures}

For each strategy, annualised excess returns and volatility are
computed over six-month and twelve-month windows immediately before
(pro-forma) and after (live) launch, using daily excess returns.
To ensure cross-strategy comparability, returns are rescaled to a common
10\% annualised volatility target:
\begin{equation}
  r^{\text{adj}}_{i,h} = r^{\text{ann}}_{i,h}
  \times \frac{\sigma^{\text{target}}}{\sigma^{\text{ann}}_{i,h}},
  \qquad \sigma^{\text{target}} = 0.10,
\label{eq:voladj}
\end{equation}
where $h \in \{6\text{m}, 12\text{m}\}$, $r^{\text{ann}}_{i,h}$
denotes the annualised excess return of strategy $i$ over window $h$,
and $\sigma^{\text{ann}}_{i,h}$ denotes the corresponding annualised
volatility of daily excess returns.\footnote{%
The vol-adjustment is \emph{in-window}: $\sigma^{\text{ann}}_{i,h}$ is
the standard deviation of daily excess returns over the same window
$h$ that is used to compute the numerator (i.e.\ the pro-forma window
for $r^{\text{adj,pre}}$ and the live window for $r^{\text{adj,live}}$),
rather than an ex-ante forecast or a rolling window extending outside
$h$.
This choice sidesteps look-ahead bias in the pro-forma period and
preserves the interpretation of $r^{\text{adj}}_{i,h}$ as the
realised Sharpe ratio over $h$ scaled to the target 10\% volatility.}
In practical terms, this normalisation places all strategies on a
common 10\% volatility footing, so that differences in performance
reflect return quality rather than variation in target risk.
Because all strategies are normalised to the same volatility,
$r^{\text{adj}}_{i,h}$ is proportional to the Sharpe ratio
($r^{\text{adj}}_{i,h} = \text{SR}_{i,h} \times \sigma^{\text{target}}$);
all performance measures are accordingly reported in
volatility-adjusted return units (per cent p.a.) throughout.

Performance decay is defined as the pro-forma-to-live change in volatility-adjusted return:
\begin{equation}
  \Delta r^{\text{adj}}_{i,h} \equiv
  r^{\text{adj,live}}_{i,h} - r^{\text{adj,pre}}_{i,h},
  \quad h \in \{6\text{m}, 12\text{m}\},
\label{eq:decay_def}
\end{equation}
Negative values indicate that live performance fell short of the
marketed backtest.

Four additional risk measures are computed over each window $h$:
maximum drawdown ($\text{MDD}_{i,h}$), conditional value-at-risk
at the 95\% confidence level (CVaR$_{95}$; \citealt{Rockafellar2002}),
downside deviation \citep{Sortino1991}, and the Sortino ratio.
For loss-magnitude metrics, risk forecast error is defined as
$\text{deterioration}_{i,h} = |\text{metric}^{\text{live}}_{i,h}|
- |\text{metric}^{\text{pre}}_{i,h}|$; positive values indicate
deterioration in live risk relative to the pro-forma period.

\subsection{Regression framework}

The levels regression tests the degree to which pro-forma performance
persists into the live period:
\begin{equation}
  r^{\text{adj,live}}_{i,h}
  = \alpha + \beta_{\text{levels}}\, r^{\text{adj,pre}}_{i,h}
  + \boldsymbol{\delta}' \mathbf{X}_i
  + \text{FE}_{b \times y}
  + \varepsilon_i,
\label{eq:levels_reg}
\end{equation}
where $r^{\text{adj,pre}}_{i,h}$ denotes the horizon-matched pro-forma
volatility-adjusted return: the six-month value for $h = 6$m and the
twelve-month value for $h = 12$m.
The control vector $\mathbf{X}_i$ is defined as
\begin{equation}
  \mathbf{X}_i = \bigl(\, r^{\text{adj,early}}_{i},\;
  \sigma^{\text{pre}}_{i,12\text{m}},\;
  \text{Age}_i \,\bigr)',
\label{eq:controls}
\end{equation}
where $r^{\text{adj,early}}_{i}$ denotes the volatility-adjusted return
computed over the pro-forma period excluding the final twelve months,
serving as a proxy for the strategy's long-run skill independent of
the immediate pre-launch window;
$\sigma^{\text{pre}}_{i,12\text{m}}$ denotes the annualised volatility
of daily returns over the twelve months preceding launch; and
$\text{Age}_i$ denotes the length of the pro-forma track record in
years.
These controls absorb variation in long-run performance history,
pro-forma risk level, and track-record maturity, respectively.
$\text{FE}_{b \times y}$ denotes bucket $\times$ launch-year fixed
effects, which absorb time-invariant bucket characteristics and common
market conditions at the launch-year level.
All continuous regressors are winsorised at the 1st and 99th
percentiles.
A coefficient $\beta_{\text{levels}} = 1$ would indicate full
portability of pro-forma performance; values below unity imply that
live expectations based on the backtest are overstated.

The corresponding decay regression replaces the dependent
variable with $\Delta r^{\text{adj}}_{i,h}$:
\begin{equation}
  \Delta r^{\text{adj}}_{i,h}
  = \alpha' + \beta_{\text{decay}}\, r^{\text{adj,pre}}_{i,h}
  + \boldsymbol{\delta}'\! \mathbf{X}_i
  + \text{FE}_{b \times y}
  + \varepsilon_i.
\label{eq:decay_reg}
\end{equation}
Because $\Delta r^{\text{adj}}_{i,h} = r^{\text{adj,live}}_{i,h}
- r^{\text{adj,pre}}_{i,h}$, the two specifications are
algebraically linked: $\beta_{\text{decay}} = \beta_{\text{levels}} - 1$.
Intuitively, $\beta_{\text{levels}} < 1$ indicates that higher
pro-forma performance predicts higher live performance, but less than
one-for-one; the corresponding negative $\beta_{\text{decay}}$
indicates that the excess does not persist.

Bucket $\times$ launch-year fixed effects are included throughout,
generating 104 non-empty cells (median cell size 12).
Standard errors are clustered at the strategy level.

The regressions above use raw volatility-adjusted returns as the
dependent variable.
The paper's central test re-estimates both specifications after
replacing the dependent variable with a benchmark-adjusted measure,
defined in Section~\ref{sec:methods_bench}, while leaving the
right-hand side unchanged.
Under the skill hypothesis---that is, if pro-forma performance
reflects strategy-specific ability beyond the common factor---$\beta_{\text{levels}}$ should remain economically large and
$\beta_{\text{decay}}$ significantly negative after benchmark
conditioning.
Under the alternative---that the backtest merely captures exposure to
a favourable bucket-factor regime---both coefficients should collapse
toward zero.

\subsection{Benchmark construction}
\label{sec:methods_bench}

The paper's central test requires a benchmark that captures the
common-factor regime shared by strategies in the same bucket without
requiring discretionary index selection.
A single market index is insufficiently granular for these strategies,
as products with similar labels frequently differ materially across
providers.
A peer-based benchmark that captures the common bucket environment
more directly is therefore employed, and the result is corroborated
with an external Bloomberg benchmark.
Agreement between two benchmarks constructed from entirely different
information sets provides robust corroboration of the main finding.

The primary benchmark is the LOO bucket-average.
For each strategy--day pair, the LOO return is the equal-weighted
average of all other strategies in the same bucket:
\begin{equation}
  \text{LOO}_{i,t} = \frac{\sum_{j \neq i,\; j \in b(i)} r_{j,t}}
                          {n_{b(i),t} - 1},
\label{eq:loo_def}
\end{equation}
where $b(i)$ denotes the bucket to which strategy $i$ is assigned and
$n_{b(i),t}$ denotes the number of strategies in that bucket with an
observed return on day $t$; the denominator $n_{b(i),t}-1$ excludes
strategy $i$ itself.
Intuitively, each strategy's benchmark is the average daily return of
every other strategy in its bucket on that day---a peer-based measure
that captures what the shared factor environment delivered without
contamination from the strategy's own return.
The summation includes all peer strategies with an observed return on
that date, regardless of whether they are in their pro-forma or live
period; the LOO benchmark therefore reflects the realised bucket-factor
regime rather than a subset conditioned on launch
status.\footnote{Restricting the LOO benchmark to only
contemporaneously live strategies yields virtually identical results;
see Appendix~\ref{app:loo_liveonly}.}
The LOO benchmark encompasses all 1,726 strategies in the analysis
sample, requires no discretionary index selection, and captures the
prevailing bucket-factor regime by construction.
Seven buckets are represented: Carry, Hedging, Momentum, Multi Premia,
Factor, Value, and Liquidity.

Equation~\ref{eq:loo_def} defines a daily benchmark.
Its annualised counterpart $\text{LOO}^{\text{ann}}_{i,h}$ is the
annualised mean of the daily LOO series over window $h$ (the first
$h$ months of live trading, compounded and expressed on an annual
basis).
The primary outcome variable is then the simple LOO-relative return,
\begin{equation}
  r^{\text{rel,LOO}}_{i,h}
  = r^{\text{ann}}_{i,h} - \text{LOO}^{\text{ann}}_{i,h},
\label{eq:loo_rel}
\end{equation}
where $r^{\text{ann}}_{i,h}$ is the strategy's own annualised return
over the same window.
That is, the peer-group average return is subtracted from the
strategy's own return over the same window; a positive value indicates
that the strategy outperformed its bucket peers.
This imposes a unit loading ($\beta = 1$), the natural null hypothesis
for strategies sharing the same return mechanism.
As a robustness check, a pre-estimated-beta variant,
$\hat\alpha^{\text{prebet}}_{i,h}$, is also reported.
This variant relaxes the unit-loading assumption by estimating each
strategy's individual loading on the LOO benchmark from the full
pro-forma history before computing live abnormal returns.\footnote{Specifically, $\beta^{\text{pre}}_i$ is
estimated from $r_{i,t} = \alpha^{\text{pre}}_i + \beta^{\text{pre}}_i
\text{LOO}_{i,t} + u_{i,t}$ over the pro-forma window
\label{eq:loo_prebet_est}
and held fixed in live:
$\text{AR}_{i,t} = r_{i,t} - \hat\beta^{\text{pre}}_i\,\text{LOO}_{i,t}$
\label{eq:loo_ar}
(annualised mean over window $h$ gives $\hat\alpha^{\text{prebet}}_{i,h}$).
Using the pro-forma window (median 5,396 trading days) reduces
$\beta$ estimation noise by $\sqrt{5396/252}\approx 4.6\times$
relative to a live-window-only estimate.}

LOO-relative decay is defined analogously:
\begin{equation}
  \Delta r^{\text{rel,LOO}}_{i,h}
  = r^{\text{rel,LOO,live}}_{i,h} - r^{\text{rel,LOO,pre}}_{i,h}.
\label{eq:loo_rel_decay}
\end{equation}

The robustness benchmark is constructed from Bloomberg total-return
indices.
Each strategy is assigned a Bloomberg index via a deterministic rule
that maps asset class to one of nine total-return indices, as detailed
in Table~\ref{tab:bloomberg_mapping}.
The mapping is applied uniformly across all ten institutions and
fixed before launch, precluding strategic benchmark selection;
coverage is complete (1,726 of 1,726 strategies).

\begin{table}[htbp]
\centering
\caption{Bloomberg Benchmark Mapping. Each strategy is assigned a
total-return index based on its asset class. Credit strategies are further
split by bucket: carry and short-volatility strategies receive the
high-yield index; all others receive the investment-grade index.
FX strategies are benchmarked against the J.P.\ Morgan USD Index,
a broad G10 currency basket.}
\label{tab:bloomberg_mapping}
\small
\begin{tabular}{lll}
\toprule
Asset class & Condition & Primary index \\
\midrule
Equities     & ---                    & MSCI ACWI              \\
Rates        & ---                    & Bloomberg US Treasury   \\
Credit       & Carry/Short-Convexity  & Bloomberg US High Yield \\
Credit       & All other buckets      & Bloomberg US IG Corp    \\
Commodities  & ---                    & Bloomberg Commodity TR  \\
Multi-Asset  & ---                    & MSCI ACWI              \\
FX           & ---                    & J.P.\ Morgan USD Index  \\
\bottomrule
\end{tabular}
\end{table}

The primary Bloomberg specification is a single-factor Jensen regression
estimated over the first $h$ months of live trading:
\begin{equation}
  r_{i,t} = \alpha_i + \beta_i\, r^{\text{bm}}_{i,t} + u_{i,t},
\label{eq:jensen}
\end{equation}
The intercept $\alpha_i$ (Jensen alpha) measures the strategy's
average daily return after removing the portion explained by its
market benchmark---that is, the return that cannot be attributed to
broad market exposure.
Because these are total-return indices, the single-factor model
captures linear (delta) exposure but not the convexity profile of
non-linear strategies such as short- and long-volatility positions
(approximately 40\% of the sample).
A two-factor variant is therefore reported that augments
Equation~\eqref{eq:jensen} with a convexity proxy based on the
VIX term-structure ratio
(CBOE 3-month implied volatility divided by 1-month implied
volatility); when the ratio exceeds unity the volatility curve is in
contango and short-volatility strategies earn carry, whereas
backwardation (ratio below unity) signals short-volatility stress:
\begin{equation}
  r_{i,t} = \alpha_i + \beta^{\text{bm}}_i\, r^{\text{bm}}_{i,t}
  + \beta^{\text{cvx}}_i\, f^{\text{cvx}}_{t} + u_{i,t},
\label{eq:jensen_convex}
\end{equation}
where $f^{\text{cvx}}_{t}$ denotes the daily change in the log
VIX term-structure ratio.
The main text reports the single-factor specification; the two-factor
variant yields identical conclusions
(Appendix~\ref{app:bloomberg_twofactor}).\footnote{When a strategy's
benchmark $R^2$ is near zero, the Jensen regression removes little
common-factor variation and $\hat\alpha$ collapses toward the raw
excess return; including such strategies therefore biases the
Bloomberg-adjusted results against the central finding. Excluding
strategies with $R^2 < 0.05$ leaves the conclusions unchanged.}
Alongside the Jensen alpha, a simple Bloomberg-relative return is
defined that imposes a unit loading on the benchmark, maintaining
comparability with the LOO simple-relative measure:
$r^{\text{rel,bm}}_{i,h} = r^{\text{ann}}_{i,h}
- r^{\text{bm,ann}}_{i,h}$,
where $r^{\text{bm,ann}}_{i,h}$ denotes the annualised mean of the
daily benchmark return $r^{\text{bm}}_{i,t}$ over window $h$.
Bloomberg-relative decay is then defined as
\begin{equation}
  \Delta r^{\text{rel,bm}}_{i,h}
  = r^{\text{rel,bm,live}}_{i,h} - r^{\text{rel,bm,pre}}_{i,h}.
\label{eq:rel_decay}
\end{equation}

\subsection{Channel variables}
\label{sec:channel_vars}

Two channel variables are introduced to capture the intensity of
regime timing and launch-cohort density (a proxy for crowding).

The first variable, regime extremity, measures the deviation of the
prevailing bucket-factor return from its long-run average at the time
of launch.
For each strategy~$i$ in bucket~$b$, it is defined as:
\begin{equation}
  \text{RegimeExtremity}_{i} \;=\;
  \bar{r}^{\,\text{LOO}}_{b(i),\,\text{pre},12\text{m}}
  \;-\;
  \mu_{b},
  \label{eq:regime_extremity}
\end{equation}
Intuitively, this is the difference between the peer-group return in
the twelve months preceding launch and the long-run bucket average.
A positive value indicates that the bucket factor was performing above
its historical average at the time of launch---what practitioners
would term a ``hot'' environment for that particular risk premium.

The regime-extremity channel is tested through three specifications.
Specification~(i) adds regime extremity as an additive control to the
baseline decay regression (Equation~\ref{eq:decay_reg}):
\begin{equation}
  \Delta r^{\text{adj}}_{i,h}
  = \alpha + \beta_1\, r^{\text{adj,pre}}_{i,h}
  + \gamma\, \text{RegimeExtremity}_{i}
  + \boldsymbol{\delta}'\! \mathbf{X}_i
  + \text{FE}_{b \times y}
  + \varepsilon_i.
\label{eq:regime_add}
\end{equation}
Specification~(ii) augments the baseline decay regression
(Equation~\ref{eq:decay_reg}) with both an additive regime-extremity
term and its interaction with the pro-forma predictor:
\begin{equation}
  \Delta r^{\text{adj}}_{i,h}
  = \alpha + \beta_1\, r^{\text{adj,pre}}_{i,h}
  + \gamma\, \text{RegimeExtremity}_{i}
  + \lambda\, r^{\text{adj,pre}}_{i,h} \times \text{RegimeExtremity}_{i}
  + \boldsymbol{\delta}'\! \mathbf{X}_i
  + \text{FE}_{b \times y}
  + \varepsilon_i.
\label{eq:regime_interact}
\end{equation}
Specification~(iii) retains the right-hand side of Specification~(i)
but replaces the dependent variable with LOO-relative decay
(Equation~\ref{eq:loo_rel_decay}), isolating the idiosyncratic
component:
\begin{equation}
  \Delta r^{\text{rel,LOO}}_{i,h}
  = \alpha + \beta_1\, r^{\text{adj,pre}}_{i,h}
  + \gamma\, \text{RegimeExtremity}_{i}
  + \boldsymbol{\delta}'\! \mathbf{X}_i
  + \text{FE}_{b \times y}
  + \varepsilon_i.
\label{eq:regime_loo}
\end{equation}

The second variable, launch density, measures the number of strategies
entering the same risk-premium bucket within a given year.
Because deployed capital is unobserved in the data, launch density
serves as a proxy for crowding pressure rather than a direct measure
of capital congestion.
For each bucket-year cell $(b, y)$, it is defined as:
\begin{equation}
  \text{LogLaunchDensity}_{b,y} \;=\;
  \ln\bigl(1 + n_{b,y}\bigr),
  \label{eq:launch_density}
\end{equation}
where $n_{b,y}$ denotes the number of strategies launched in bucket~$b$
during year~$y$.
A high value indicates that numerous strategies entered the same
risk-premium bucket during the same year, consistent with multiple
institutions simultaneously targeting the same premium.
The mean launch density is 17.8 strategies per bucket-year (median~10,
range 1--105).

Density is constant within bucket-year cells, so its identifying
variation decomposes into three components: cross-bucket level
differences, cross-year level differences (the industry-wide launch
cycle), and bucket-year residuals.
All three specifications include bucket fixed effects only.
Bucket fixed effects guard against confounding between density and
structural differences between product categories: smaller
categories have few providers and distinct payoff dynamics, while
larger categories draw many providers and have their own return
profile, and a non-zero density coefficient estimated without
bucket fixed effects could not be distinguished from these
bucket-composition effects.
Launch-year fixed effects are omitted so that the industry-wide
entry cycle---a substantial component of the crowding
mechanism---remains in the identifying variation; adding year
fixed effects would identify the density coefficient only from
within-year cross-bucket residuals.

The launch-density channel is tested through three specifications.
Specification~(A) adds log launch density as an additive control to
the baseline decay regression (Equation~\ref{eq:decay_reg}), with
bucket fixed effects in place of the bucket~$\times$~launch-year
interaction to preserve the entry-cycle variation:
\begin{equation}
  \Delta r^{\text{adj}}_{i,h}
  = \alpha + \beta_1\, r^{\text{adj,pre}}_{i,h}
  + \gamma\, \text{LogLaunchDensity}_{b,y}
  + \boldsymbol{\delta}'\! \mathbf{X}_i
  + \text{FE}_{b}
  + \varepsilon_i.
\label{eq:density_add}
\end{equation}
Specification~(B) augments the baseline decay regression
(Equation~\ref{eq:decay_reg}) with both the additive density term and
its interaction with the pro-forma predictor:
\begin{equation}
  \Delta r^{\text{adj}}_{i,h}
  = \alpha + \beta_1\, r^{\text{adj,pre}}_{i,h}
  + \gamma\, \text{LogLaunchDensity}_{b,y}
  + \lambda\, r^{\text{adj,pre}}_{i,h} \times \text{LogLaunchDensity}_{b,y}
  + \boldsymbol{\delta}'\! \mathbf{X}_i
  + \text{FE}_{b}
  + \varepsilon_i.
\label{eq:density_interact}
\end{equation}
Specification~(C) retains the right-hand side of Specification~(A)
but replaces the dependent variable with LOO-relative decay
(Equation~\ref{eq:loo_rel_decay}), stripping out the common bucket
factor and isolating strategy-specific deterioration (directly
analogous to Specification~(iii) of the regime-extremity table):
\begin{equation}
  \Delta r^{\text{rel,LOO}}_{i,h}
  = \alpha + \beta_1\, r^{\text{adj,pre}}_{i,h}
  + \gamma\, \text{LogLaunchDensity}_{b,y}
  + \boldsymbol{\delta}'\! \mathbf{X}_i
  + \text{FE}_{b}
  + \varepsilon_i.
\label{eq:density_loo}
\end{equation}

\section{Empirical Results}
\label{sec:results}

This section develops the paper's central argument in four steps.
The first establishes that the raw pro-forma-to-live gap is
economically large and exceeds what mechanical mean reversion would
predict.
The second adjusts for within-bucket peer performance---the paper's
main result.
The third confirms robustness using an independent Bloomberg benchmark.
The fourth investigates why backtests mislead, identifying regime
extremity as the primary channel and launch density as a secondary
proxy for crowding.

Table~\ref{tab:pf_vs_live_a} reports mean volatility-adjusted returns
in the pro-forma and live windows, together with the decay (live minus
pro-forma), for the full sample and by asset class.
Results are reported at both six-month and twelve-month post-launch
horizons; significance stars denote rejection of the null hypothesis
of zero mean decay.
The central finding is strikingly consistent: nearly every asset class
exhibits negative mean performance decay at both horizons.

  \begingroup
  \singlespacing
  \scriptsize
  \setlength{\tabcolsep}{3pt}%
  \renewcommand{\arraystretch}{0.90}%
  \input{table2_pf_vs_live_a}  \endgroup

\TableBarrier

For the full sample, the mean volatility-adjusted return in the six
months preceding launch is 3.6\% p.a., compared with 1.5\% in
the live period---a decline of 2.1 percentage points ($p < 0.01$).
Over the twelve-month horizon, the gap widens: pro-forma performance
of 4.1\% declines to 1.0\% in the live period, a decay of $-3.1$
percentage points ($p < 0.01$).
Decay is also heterogeneous across asset classes;
Table~\ref{tab:pf_vs_live_a} reports the breakdown.
Among the broader portfolio roles, Hedging/Defensive strategies
exhibit comparatively smaller raw decay, a pattern
Section~\ref{sec:implications} notes is consistent with their
distinct payoff role but does not test directly.

Event-time plots centred on the launch date confirm a distinct
structural break: performance that persists through the pre-launch
period deteriorates sharply in the months following launch
(Appendix~\ref{app:event_time}).

Part of the observed decay is mechanical: any mean-reverting series
will exhibit apparent deterioration when conditioned on a high-water
mark.
Supplementary tests indicate that the launch boundary contains
information beyond mechanical mean reversion.
A moving-block bootstrap (Appendix~\ref{app:bootstrap_sensitivity})
demonstrates that the twelve-month decay coefficient exceeds what
sampling noise alone can produce ($p < 0.001$, robust across block
sizes of 10--63 trading days).
A per-strategy placebo test (Appendix~\ref{app:placebo}) further
confirms that launch months are not systematically timed to each
strategy's own pre-launch peaks on the vol-adjusted return signal,
ruling out within-strategy cherry-picking as an explanation for the
observed decay.

\subsection{Pro-forma predictability under benchmark adjustment}
\label{sec:loo_results}

The key question for allocators is not whether raw backtests and
live outcomes are correlated.
It is whether that association survives once performance is measured
against an appropriate benchmark.
The central test replaces the raw return dependent variable with
benchmark-adjusted live performance under the LOO and Bloomberg
benchmarks defined in Section~\ref{sec:data}.

After benchmark adjustment, the median strategy underperforms its LOO
bucket peers by approximately 0.8 percentage points p.a.\ at the
twelve-month horizon (0.7 percentage points at six months), and 59\%
of strategies
deliver negative LOO-relative returns (55\% at six months).
Measured against the external Bloomberg total-return benchmark, which
uses an information set entirely outside the strategy universe, the
median underperformance widens to 3.0 percentage points at twelve
months (2.8 percentage points at six months), with 59\% of strategies
negative at
twelve months (56\% at six months).
The full distribution under both benchmarks is reported in
Table~\ref{tab:bench_rel_dist} (Appendix~\ref{app:bench_rel_dist}).

Table~\ref{tab:bench_summary_levels} presents the core result.
Each row reports the OLS coefficient on the key pro-forma predictor
(horizon-matched volatility-adjusted return) when the dependent
variable is live performance measured under a different benchmark
specification.
In raw returns, marketed backtests appear informative.
Once live outcomes are benchmarked appropriately, however, most of
that apparent information disappears.

\input{table_benchmark_summary_levels}
\TableBarrier

Without benchmark adjustment, the horizon-matched pro-forma predictor
yields $\hat\beta = 0.137$ ($p < 0.001$, $R^2 = 0.148$).
This positive partial correlation confirms that pro-forma performance
carries cross-sectional information about which strategies will
perform better in the live period---but only in the unconditional,
unadjusted sense.

When the dependent variable is replaced by the OLS-estimated Jensen
$\alpha$ against each strategy's assigned Bloomberg total-return
index, the coefficient declines to $0.025$ ($p < 0.05$).
The $R^2$ simultaneously declines from 0.148 to 0.032.
Although the coefficient retains marginal statistical significance,
its economic magnitude is sharply reduced---a decline of approximately
81\% relative to the raw estimate---indicating that the marketed
pro-forma signal is predominantly driven by common-factor exposure
rather than portable alpha.

Under the LOO peer-average benchmark, the coefficient is
$0.034$ ($p < 0.01$, $R^2 = 0.054$), a reduction of 75\% relative
to the raw estimate.
The close agreement between the peer-based and external benchmarks is
noteworthy.
These two benchmarks draw on entirely different information---the LOO
benchmark uses only peer-strategy returns, while the Bloomberg
benchmark uses broad market indices---yet both remove the majority of
the raw signal.
This convergence makes it unlikely that the attenuation is an artefact
of a particular benchmark construction choice, and the external
Bloomberg benchmark also rules out LOO endogeneity---the concern that
institutions clustering launches into favourable regimes might
contaminate the peer average.\footnote{The median $R^2$ from
regressing each strategy's daily returns on its LOO benchmark over the
full pro-forma window is 0.074, confirming that the LOO factor
captures a meaningful share of daily return variation and is not so
noisy as to mechanically drive the coefficient toward zero.}

Table~\ref{tab:bench_summary_decay} turns to the decay outcome: the
change in benchmark-relative return from the pro-forma to the live
window (Equation~\ref{eq:loo_rel_decay} for LOO,
Equation~\ref{eq:rel_decay} for Bloomberg).
The mean LOO-relative decay is $-1.3$\% p.a.\ at twelve months,
with 56\% of strategies exhibiting negative decay.

\input{table_benchmark_summary_decay}
\TableBarrier

In absolute terms, the horizon-matched pro-forma predictor yields
$\hat\beta = -0.896$ ($p < 0.001$, $R^2 = 0.494$).
This large $R^2$ is partly mechanical---decay is defined as live minus
pro-forma, so higher pro-forma returns mechanically generate larger
measured decay if live returns mean-revert---but the bootstrap results
reported above confirm that the magnitude exceeds what pure mean
reversion would produce.

Benchmark-relative decay remains economically meaningful but is
substantially smaller than the raw decay estimate.
When decay is measured relative to the Bloomberg benchmark---removing
the common-factor return from both the pro-forma and live
windows---the coefficient declines to $-0.377$ ($p < 0.001$,
$R^2 = 0.247$), a reduction of approximately 58\%.
Under the LOO benchmark, the decay coefficient is $-0.372$
($p < 0.001$, $R^2 = 0.267$), virtually identical in magnitude.
The common-factor environment therefore accounts for the majority of
the marketed backtest signal, although a smaller residual component
persists after benchmark adjustment.
The close similarity of the two adjusted coefficients ($-0.377$ vs.\
$-0.372$) suggests that both benchmarks remove the same broad
systematic component; a two-factor convexity variant yields the same
result (Appendix~\ref{app:bloomberg_twofactor}).\footnote{The Bloomberg
benchmark is conservative for the approximately 40\% of strategies
with low benchmark $R^2$ or convex payoff profiles, where Jensen alpha
approximates the raw return; this conservatism works against the
paper's central finding.}
A natural concern is that benchmark adjustment simply adds noise to
the dependent variable, mechanically attenuating the coefficient.
Two tests argue against this: on a matched sample, a Wald test
comparing the raw and adjusted coefficients rejects equality
($z = 4.03$, $p < 0.001$; Appendix~\ref{app:raw_vs_adjusted}),
confirming that the attenuation exceeds what added noise alone would
produce; and including the benchmark return as a right-hand-side
control leaves the raw coefficient virtually unchanged, so the
attenuation appears only when performance is measured relative to the
benchmark---a pattern consistent with genuine common-factor removal
rather than noise inflation.
Benchmark design is therefore substantive, not cosmetic, in allocator
due diligence.

The implication is not that backtests are uninformative, but that
the majority of what they capture is the common-factor environment
into which the strategy was launched.
The remainder of this section examines why, focusing on regime
extremity as the primary channel and launch density as a secondary
proxy for crowding.

\subsection{Regime extremity}
\label{sec:regime_extremity}

The regime-timing mechanism predicts that strategies launched
after unusually strong bucket-factor performance should
experience greater performance decay, because the pro-forma window
over-weights a favourable factor environment that subsequently
mean-reverts.
Table~\ref{tab:regime_extremity} tests this prediction using three
specification families (Equations~\ref{eq:regime_add}--\ref{eq:regime_loo}),
each estimated at both the six-month and twelve-month horizons.

Specification~(i) (Equation~\ref{eq:regime_add}) adds regime extremity
as an additive control to the baseline raw-decay regression.
Specification~(ii) (Equation~\ref{eq:regime_interact}) includes the
interaction of pro-forma return with regime extremity, testing whether
the pro-forma signal is amplified when the bucket regime is more
extreme.
Specification~(iii) (Equation~\ref{eq:regime_loo}) is the most
informative: it replaces the dependent variable with LOO-relative
decay, removing the common bucket factor so that the coefficient on
regime extremity captures only the idiosyncratic overstatement---the
component that cannot be attributed to the shared factor environment.

\input{table_regime_extremity_paper}
\TableBarrier

In raw-decay terms, regime extremity is largely absorbed by the bucket
$\times$ launch-year fixed effects, which already capture much of the
common-factor variation.
The more informative test is Specification~(iii), which removes the
common factor from the dependent variable via the LOO benchmark.
Under this specification, regime extremity is highly significant:
$\hat\beta = 0.716$ (SE~$= 0.124$, $p < 0.001$) at the twelve-month
horizon and $\hat\beta = 0.469$ (SE~$= 0.176$, $p < 0.01$) at six
months.
Because the variable is signed, with positive values denoting hot
regimes (bucket return in the pre-launch window above its long-run
mean) and negative values denoting cold regimes (bucket return below
its long-run mean), the positive coefficient indicates that strategies
launched in hot regimes experience worse LOO-relative decay, whereas
those launched in cold regimes may experience a slight improvement.
The effect is directional rather than symmetric: hotter pre-launch
regimes inflate the pro-forma record relative to the strategy's true
peer standing, and this overstatement reverses as the regime
mean-reverts after launch.
The $R^2$ of 0.293 at the twelve-month horizon is lower than the
0.494 for raw return decay, reflecting the greater noise in the
LOO-residual outcome, but regime extremity alone contributes
meaningful explanatory power within this more demanding specification.
Restricting the LOO benchmark to contemporaneously live strategies
and applying a LOO bucket mean strengthens these estimates
(Appendix~\ref{app:loo_liveonly}), confirming that the baseline
specification is conservative.
A pre-estimated-beta specification that absorbs cross-sectional
variation in factor loadings yields a qualitatively identical
regime-extremity coefficient, ruling out loading heterogeneity as
a mechanical driver.

\begin{figure}[htbp]
  \centering
  \includegraphics[width=0.75\textwidth]{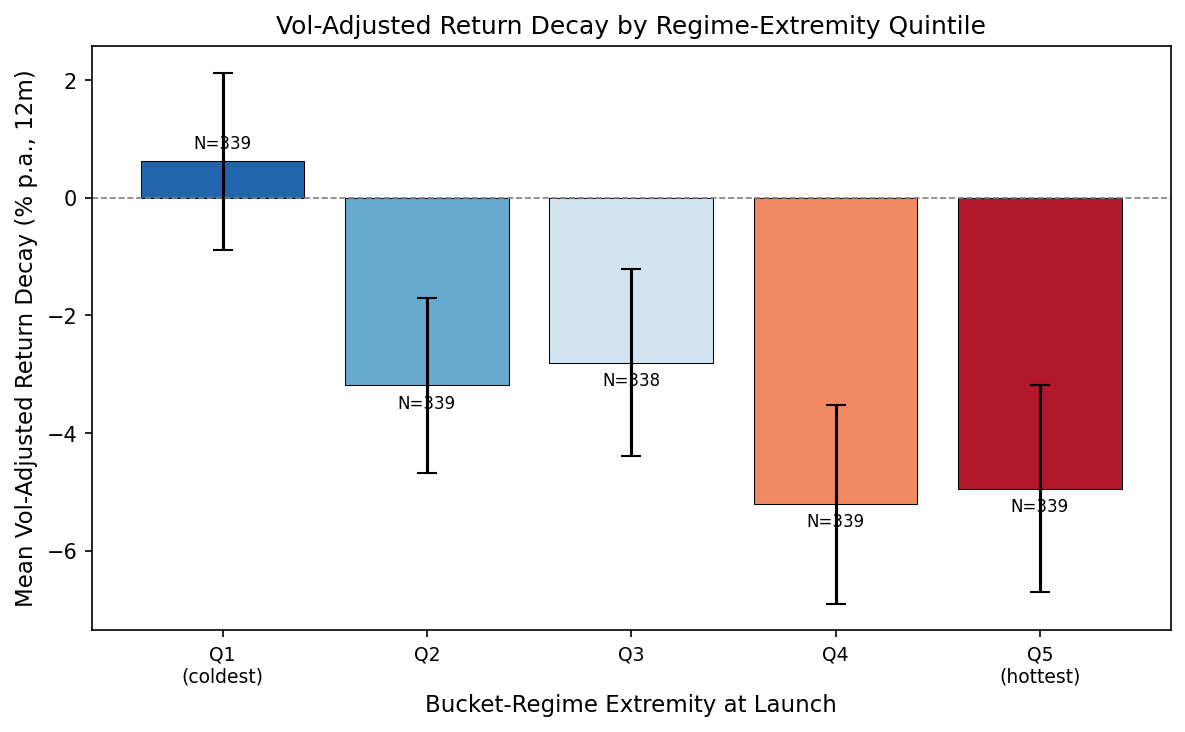}
  \caption{Mean Return Decay by Regime-Extremity Quintile.
           Strategies are sorted into quintiles of regime extremity
           (Equation~\ref{eq:regime_extremity}), from Q1 (coldest
           bucket-factor regime at launch) to Q5 (hottest).
           Bars show mean twelve-month return decay
           ($\Delta r^{\text{adj}}_{i,12\text{m}}$) within each bin;
           error bars are 95\% confidence intervals based on
           within-bin standard errors.
           $N \approx 339$ per bin.
           The overall gradient---from a slight improvement in Q1 to
           decay exceeding $-3$\% in Q4 and Q5---confirms that
           strategies launched after unusually strong bucket-factor
           performance experience sharper decay.
           This figure is intended as a raw-data illustration of the
           regime-timing pattern; the formal test remains the
           LOO-relative regression reported in
           Table~\ref{tab:regime_extremity}.}
  \label{fig:regime_quintiles}
\end{figure}

Figure~\ref{fig:regime_quintiles} provides a non-parametric
illustration of the same relationship.
The 1,694 strategies with available twelve-month data are sorted into
quintiles of regime extremity, and the mean twelve-month return decay
within each bin is plotted.
The pattern is monotonically increasing in magnitude from cold to hot
regimes: strategies launched in the coldest regime quintile (Q1)
experience mean decay of approximately $+0.8$ percentage points (a
slight improvement), whereas those in the hottest quintiles (Q4 and
Q5) experience decay of approximately $-3.5$ to $-4.5$ percentage
points.
The 95\% confidence intervals are wide for individual quintiles,
particularly Q4 and Q5, but the overall gradient from positive decay
in Q1 to sharply negative decay in Q5 is consistent with the
regression evidence in Table~\ref{tab:regime_extremity}.

\subsection{Launch density}
\label{sec:launch_density}

Launch density---the log number of competing strategy launches in
each bucket-year---is examined as a proxy for crowding
(Equations~\ref{eq:density_add}--\ref{eq:density_loo}).
Table~\ref{tab:launch_density} reports the results.
Specification~(A) (Equation~\ref{eq:density_add}) adds log launch
density as an additive control to the baseline raw-decay regression,
and Specification~(B) (Equation~\ref{eq:density_interact}) augments
this with the interaction of pro-forma return and log launch density.
The evidence does not support a direct crowding effect on average
decay: the coefficient on log launch density is effectively zero at
both horizons.
Specification~(C) (Equation~\ref{eq:density_loo}) sharpens this test
by replacing the dependent variable with LOO-relative decay, so that
common-factor variation is purged and any remaining coefficient
isolates strategy-specific underperformance relative to peers in the
same bucket.
The density coefficient is again statistically indistinguishable
from zero at both horizons, indicating that denser launch cohorts do
not systematically underperform their bucket peers---the
individual-strategy capacity-erosion prediction of the crowding
literature is not supported in this sample.
Crowded environments do, however, amplify the pro-forma signal at
short horizons.
At the six-month horizon, the interaction of pro-forma return with
log launch density is positive and significant, although the implied
economic magnitude is modest.
The horizon asymmetry---significant at six months but not at
twelve---is consistent with short-lived capacity erosion that
dissipates as positions are unwound.

\input{table_launch_density_paper}
\TableBarrier

\section{Implications for Due Diligence and Portfolio Construction}
\label{sec:implications}

What do these findings imply for practice?
This section draws two practical implications.
First, risk metrics deteriorate in tandem with returns, so position
sizes calibrated on pro-forma statistics are likely too aggressive.
Second, Hedging/Defensive strategies require a payoff-sensitive
evaluation framework distinct from unconditional return comparisons.

\subsection{Risk deterioration and position sizing}
\label{sec:risk}

If pro-forma returns overstate live performance because of
regime-dependent selection---as the benchmark-conditioning and
regime-timing results jointly suggest---the same mechanism should
inflate pro-forma risk metrics.
Each metric is computed over matched 12-month windows immediately
before and after launch, strategy by strategy, and a strategy is
counted as \textit{worse live} when the live value is higher than the
pro-forma value for the four dispersion and loss measures (annualised
volatility, maximum drawdown, downside deviation, CVaR$_{95}$) or
lower than the pro-forma value for the reward-to-risk ratio (Sortino).
Table~\ref{tab:risk_summary} shows that risk deterioration is common,
particularly for tail- and downside-sensitive measures.
Across five standard risk metrics, the proportion of strategies with
worse live values ranges from 49\% to 59\%.
Two-sided binomial tests against the 50\% null reject at the 1\% level
for maximum drawdown (54\%, $p = 0.002$) and Sortino ratio
(59\%, $p < 0.001$); the three other metrics are not distinguishable
from chance.

\begin{table}[htbp]
\caption{Risk Deterioration Summary.
         Fraction of strategies for which the live value of each risk metric
         is worse than the corresponding pro-forma value, computed over
         matched 12-month windows immediately before and after launch.
         Annualised volatility: standard deviation of daily returns
         $\times \sqrt{252}$.
         Maximum drawdown: largest peak-to-trough decline in
         cumulative NAV.
         Downside deviation: root-mean-square of negative daily
         returns.
         CVaR$_{95}$: expected daily loss beyond the 95th percentile
         (Expected Shortfall).
         Sortino ratio: annualised return divided by downside
         deviation.
         The $p$ column reports a two-sided binomial $p$-value
         against the null hypothesis that the fraction equals 50\%;
         $^{***}\,p<0.01$, $^{**}\,p<0.05$, $^{*}\,p<0.10$.}
\label{tab:risk_summary}
\begin{tabular}{l c c}
\toprule
Risk metric & \% strategies worse live & $p$ \\
\midrule
Annualised volatility            & 49\% & 0.395 \\
Maximum drawdown                 & 54\% & 0.002\textsuperscript{***} \\
Downside deviation               & 50\% & 0.754 \\
CVaR$_{95}$ (Expected Shortfall) & 51\% & 0.400 \\
Sortino ratio                    & 59\% & $<$0.001\textsuperscript{***} \\
\bottomrule
\end{tabular}
\end{table}
\TableBarrier

The practical implication extends beyond return overstatement: risk
also deteriorates after launch.
Position sizes calibrated on pro-forma volatility and downside
statistics are therefore likely too aggressive; allocators may
simultaneously overestimate expected return and underestimate the risk
budget required to hold the strategy.
The marketed backtest should therefore be discounted not only on
expected-performance grounds but also on position-sizing and
downside-risk grounds.

\subsection{A regime-conditional haircut rule}
\label{sec:haircut_rule}

The benchmark-conditioning and regime-extremity results together
suggest a practical rule for discounting marketed backtests.
Expected live return is modelled as a linear function of the pro-forma
statistic and the pre-launch regime:
\begin{equation}
  \label{eq:haircut}
  \widehat{r}^{\,\text{live}}_{12m}
    \;=\;
  \underbrace{\hat{\lambda}_{0}}_{\approx\,0}
    \;+\;
  \underbrace{\hat{\lambda}_{1}}_{\approx\,0.137}
    \cdot r^{\,\text{pf}}_{12m}
    \;-\;
  \underbrace{\hat{\gamma}}_{\approx\,5}
    \cdot \widetilde{x}^{\,\text{reg}}_{bt},
\end{equation}
where $r^{\,\text{pf}}_{12m}$ is the pro-forma vol-adjusted return
(percentage points p.a.) and $\widetilde{x}^{\,\text{reg}}_{bt}$ is the bucket-factor
regime-extremity z-score at launch (the standardised deviation of the
bucket factor from its long-run mean in the 12 months before launch).
$\hat{\lambda}_{1} \approx 0.137$ is the raw cross-sectional slope of
live on pro-forma vol-adjusted return reported in Row~1 of
Table~\ref{tab:bench_summary_levels}.
$\hat{\gamma} \approx 5$ percentage points per unit $z$ is a rounded calibration
chosen to reproduce the four-percentage-point cold-minus-hot quintile
spread in twelve-month decay documented in
Section~\ref{sec:regime_extremity} (Figure~\ref{fig:regime_quintiles}):
each one-standard-deviation increase in pre-launch regime extremity
warrants an additional ${\sim}5$ percentage points downward adjustment to the
expected live return.

As a worked example, consider an FX-carry strategy with a twelve-month
pro-forma vol-adjusted return of $+12$ percentage points at launch, bucketed into the
FX carry peer group, and suppose the G10 carry bucket-factor z-score
at launch is $\widetilde{x}^{\,\text{reg}}=+1.5$ (a hot regime,
roughly the top 15\% of historical observations).
Applying Equation~\ref{eq:haircut},
\[
  \widehat{r}^{\,\text{live}}_{12m}
  \;=\;
  0 \;+\; 0.137 \times 12 \;-\; 5 \times 1.5
  \;=\;
  1.64 \;-\; 7.5 \;=\; -5.86 \text{ percentage points p.a.}
\]
The pro-forma return of $+12$ percentage points p.a.\ should therefore be discounted
to an expected live return of roughly $-6$ percentage points p.a.\ once the
cross-sectional slope and regime extremity are accounted for.
This is substantially more conservative than the ad-hoc 50\% haircut
commonly applied in practice, and is consistent with the four-plus-percentage-point
cold-minus-hot quintile spread documented in
Section~\ref{sec:regime_extremity}.

\subsection{Out-of-sample failure prediction}
\label{sec:classifier_oos}

Pro-forma characteristics also support direct classification of
strategies likely to fail in live trading.
For each objective group $g$ and failure outcome $k$ we model the
probability of failure as
\begin{equation}
  \label{eq:failure_classifier}
  \Pr\!\left(y_{i,k}=1 \mid \mathbf{x}_i\right)
    \;=\;
  F\!\left(\alpha_{g,k} + \mathbf{x}_i^{\top}\boldsymbol{\beta}_{g,k}\right),
\end{equation}
where $y_{i,k}\in\{0,1\}$ takes the value one if strategy $i$ fails
in live trading under criterion $k$ and zero otherwise;
$\mathbf{x}_i$ is the vector of pro-forma predictors used in the
decay regressions (pre-launch vol-adjusted return, volatility,
maximum drawdown, CVaR$_{95}$, downside deviation, strategy age at
launch, complexity, and commonality), augmented with bucket and
launch-year fixed effects; $\alpha_{g,k}$ is a group- and
outcome-specific intercept; and $\boldsymbol{\beta}_{g,k}$ is the
corresponding coefficient vector.
The link function $F(\cdot)$ is the identity for the linear
probability model and the logistic cumulative distribution function
(CDF), $F(z) = 1/(1 + e^{-z})$, for the logit.
Coefficients are estimated separately within each $(g,k)$ cell.

Five failure criteria are considered, each evaluated over the first
12 live months.
Negative return flags strategies whose realised live Sharpe ratio is
below zero.
The three elemental risk/decay criteria use within-group
cross-sectional quantiles as data-driven thresholds: Performance
Decay flags strategies whose pro-forma-to-live vol-adjusted return
decay falls in the worst quartile of their objective group (below
the 25th percentile of $\Delta_i$); Severe Drawdown flags those
whose live maximum drawdown is in the worst quartile; and Tail Loss
flags those whose live CVaR$_{95}$ is in the worst quartile (above
the 75th percentile of this loss measure).
Multiple Failures is a composite indicator equal to one when at
least two of the four elemental criteria are triggered
simultaneously.
Within-group thresholds are re-estimated on the training fold only
to prevent information leakage.

The validation reported in Table~\ref{tab:classifier_oos} is a
time-split design: the model is trained on strategies launched
before the median launch year and tested on later-launch cohorts.
Model performance is summarised by the area under the receiver
operating characteristic curve (AUC), which ranks predictions across
the full range of classification thresholds: the receiver operating
characteristic (ROC) curve traces the trade-off between the
true-positive rate and the false-positive rate as the threshold is
swept from zero to one, and its area equals the probability that a
randomly chosen failing strategy receives a higher predicted risk
score than a randomly chosen non-failing one (0.5 = random ranking,
1.0 = perfect separation).

\input{table_r3_failure_oos_compact}
\TableBarrier

The strongest discrimination is obtained for decay- and
multiple-failure outcomes in the Return-Seeking group: AUC is 0.81
for performance decay and 0.78 for multiple concurrent failures,
with top-decile capture rates of 29\% and 20\% respectively,
implying 2--3$\times$ lift over the unconditional base rate.
Discrimination is weaker for the remaining objective groups and
outcomes (AUCs of 0.55--0.64 and lifts of 1.2--1.4$\times$), and
the weakest results arise for unconditional negative-return outcomes
in the Return-Seeking and Carry/Short-Convexity groups (AUCs of
0.48 and 0.55, respectively).
The asymmetry is consistent with the rest of the paper:
negative-return outcomes are harder to predict than decay outcomes
because the former absorb both strategy-specific and benchmark-level
noise.

As a concrete operationalisation of the haircut rule in
Equation~\ref{eq:haircut}, an allocator could combine the
regime-conditional expected-return estimate with a top-decile
exclusion filter drawn from the decay classifier: strategies with
pro-forma characteristics placing them in the top 10\% of predicted
decay risk would be excluded or heavily haircut, which empirically
captures 29\% of realised decay failures in the Return-Seeking group
at a 3$\times$ lift over the baseline fail rate.

An interpretive caveat applies to strategies whose intended role is
not unconditional return generation.
Hedging/Defensive strategies are typically held for their payoff in
stress states rather than for standalone carry, and Carry/Short-Convexity
strategies have concave payoff profiles whose risk is poorly summarised
by volatility-adjusted return alone.
The decay estimates reported above measure pro-forma-to-live change in
unconditional vol-adjusted return for all strategies on the same
footing, which is the correct measure for testing whether the marketed
backtest signal persists.
It is not, however, the appropriate criterion for judging whether
Hedging/Defensive strategies are fulfilling their intended portfolio
function: a state-dependent payoff evaluation conditioned on equity
drawdowns or volatility spikes would be required for that judgment,
and is left for future work.
The smaller raw decay observed for the Hedging/Defensive group should
therefore not be over-interpreted as evidence of superior portability,
and the present analysis takes no stand on whether these strategies
deliver their intended insurance function in live trading.

\section{Conclusion}
\label{sec:conclusion}

This paper examines how institutional allocators should interpret
marketed backtests of structured investment strategies.
The analysis contributes in three ways.
First, it quantifies the gap between pro-forma and live performance on
a uniquely large commercial sample of 1,726 strategies from ten global
institutions over 2009--2025.
Second, it shows that once live performance is measured against a
leave-one-out bucket-average peer benchmark, the residual information
content of the marketed backtest is economically negligible: what looks
like strategy-specific skill is predominantly the common factor regime
prevailing at launch.
Third, it identifies two structural channels---regime timing at launch
and a horizon-dependent launch-density effect---that jointly explain
the residual decay, and translates the result into an operational
rule: the haircut applied to a marketed backtest should increase with
the extremity of the pre-launch factor regime.

For allocators, the practical implications are threefold.
Backtests should be benchmarked against appropriate peers, positions
should be sized more conservatively than pro-forma risk metrics
suggest, and hedging strategies should be evaluated according to their
payoff role rather than unconditional carry.

Three limitations warrant discussion.
The sample is positively selected because strategies proposed but
never launched are unobserved; the proprietary dataset limits exact
replication; and inference based on ten institution clusters should
be interpreted with caution.

Two directions for future research follow naturally.
Access to rejected strategies would sharpen the selection-bias bound,
and tracking bucket-relative alpha over multi-year live horizons would
clarify whether the regime-timing decay documented here is a transient
launch artefact or a persistent drag on performance.

\section*{Declarations}

\paragraph{Funding.}
No external funding was received for this research.

\paragraph{Conflict of interest / Competing interests.}
The author is employed by Resonanz Capital GmbH, an institutional asset
manager that provided access to the proprietary strategy-level dataset
analysed in this paper.
Resonanz Capital has no financial interest in the specific strategies
studied, did not participate in the design of the empirical tests, the
data analysis, or the drafting of this manuscript, and did not review
the findings prior to submission.
The author declares no other competing interests.

\paragraph{Data availability.}
The underlying strategy-level data were provided under a commercial
licence by Resonanz Capital GmbH and cannot be redistributed.
Aggregated descriptive statistics, regression output, and all
intermediate tables reproduced in the paper are available from the
author on reasonable request.
Documentation of the data construction pipeline is available from the
author for the purpose of replication review.

\paragraph{Code availability.}
The code used to construct the analytic sample and estimate all
regressions is available from the author on reasonable request, subject
to the restrictions on redistributing the underlying strategy-level
data noted above.

\paragraph{Ethics approval.}
Not applicable.

\paragraph{Author contributions.}
The author is the sole contributor and is responsible for the
conception, data preparation, analysis, and writing of this paper.

\newpage
\bibliographystyle{apalike}

\newpage

The appendices below supplement the main text.

\begin{appendices}
\makeatletter
\renewcommand{\@Alph}[1]{\AlphAlph{#1}}%
\makeatother
\renewcommand{\thefigure}{\thesection.\arabic{figure}}
\renewcommand{\thetable}{\thesection.\arabic{table}}
\counterwithin{figure}{section}
\counterwithin{table}{section}

\section{Sample Detail: Institution and Launch-Year Composition}
\label{app:sample_detail}

Table~\ref{tab:sample_detail} reports the breakdown of the
$N = 1{,}726$ live strategies in the baseline analysis sample by
originating institution and by launch-year cohort.  Institutions
are anonymised as A--J.  The launch-year cohort 2025 is
partial (it ends at the sample cutoff date) and therefore
contributes a smaller share than a full annual cohort would.

\begin{table}[htbp]
\centering
\caption{Sample Detail: Institution and Launch-Year Composition.
  Number of live strategies and share of the baseline sample
  ($N = 1{,}726$) by originating institution (columns~1--3) and by
  launch-year cohort (columns~4--6).  Institution identifiers are
  anonymised.  The 2025 cohort is partial, ending at the sample
  cutoff date.}
\label{tab:sample_detail}
\small
\begin{tabular}{l r r @{\hspace{2em}} l r r}
\toprule
Group & N & Share & Group & N & Share \\
\midrule
{By institution} &  &  & {By launch year} &  &  \\
\quad Inst.~A & 208 & 12\% & \quad 2009--2015 & 152 & 9\% \\
\quad Inst.~B & 353 & 20\% & \quad 2016--2018 & 358 & 21\% \\
\quad Inst.~C & 193 & 11\% & \quad 2019--2021 & 606 & 35\% \\
\quad Inst.~D & 108 & 6\% & \quad 2022--2024 & 524 & 30\% \\
\quad Inst.~E & 392 & 23\% & \quad 2025 & 86 & 5\% \\
\quad Inst.~F & 93 & 5\% &  &  &  \\
\quad Inst.~G & 21 & 1\% &  &  &  \\
\quad Inst.~H & 154 & 9\% &  &  &  \\
\quad Inst.~I & 184 & 11\% &  &  &  \\
\quad Inst.~J & 20 & 1\% &  &  &  \\
\bottomrule
\end{tabular}
\end{table}
\TableBarrier

\section{Event-Time Performance Around Launch}
\label{app:event_time}

\begin{figure}[htbp]
  \centering
  \begin{subfigure}[t]{0.48\textwidth}
    \centering
    \includegraphics[width=\textwidth]{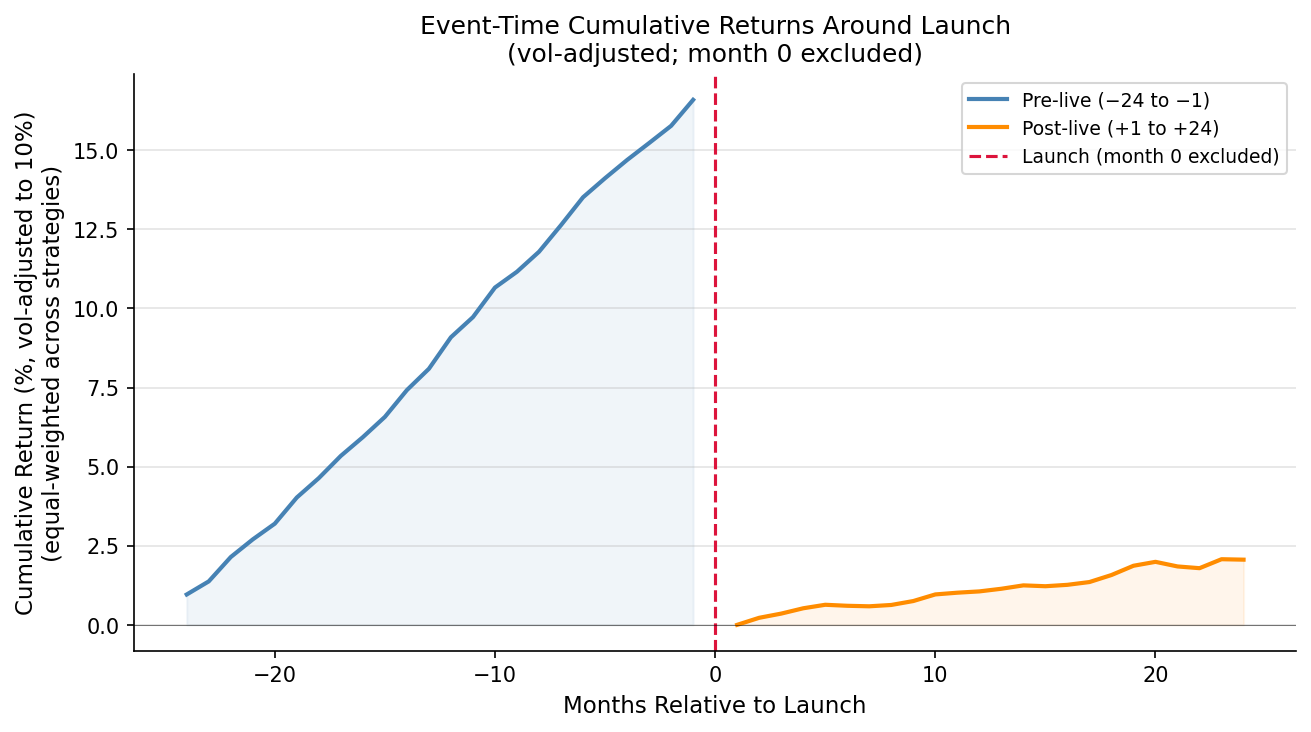}
    \caption{Cumulative mean excess return (event time)}
    \label{fig:event_returns}
  \end{subfigure}
  \hfill
  \begin{subfigure}[t]{0.48\textwidth}
    \centering
    \includegraphics[width=\textwidth]{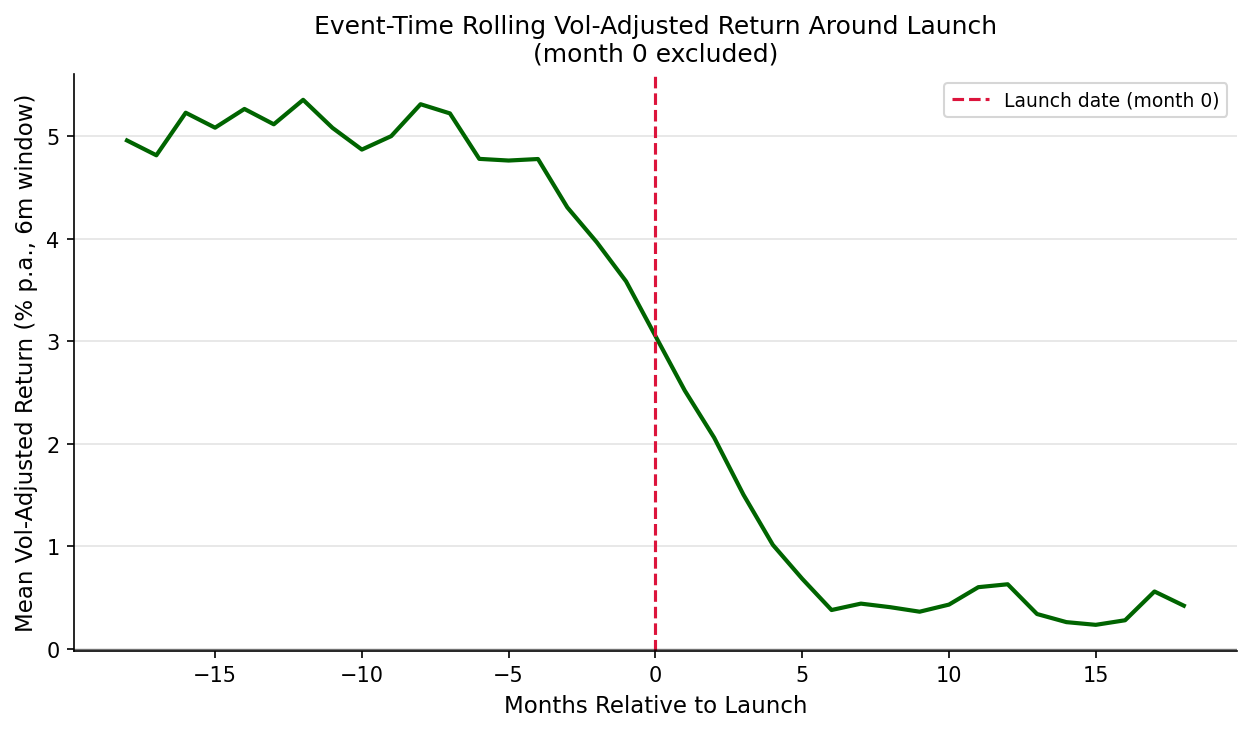}
    \caption{Event-time rolling vol-adjusted return}
    \label{fig:event_sharpe}
  \end{subfigure}
  \caption{Event-Time Performance Around Strategy Launch.
           Panel~(a): cumulative equal-weighted mean excess return
           (vol-adjusted to 10\% annualised target) in event time
           relative to launch (month~0).
           Panel~(b): rolling six-month vol-adjusted return in event time.
           The rolling mean return falls from approximately $+5\%$~p.a.\
           in the six months before launch to approximately
           $+0.5\%$~p.a.\ in the six months after, a drop of roughly
           $4.5$~percentage points p.a.\ concentrated around the launch
           date.  The step-down at month~0 is the visual manifestation
           of the pro-forma-to-live decay documented in
           Section~\ref{sec:results}.}
  \label{fig:event_time}
\end{figure}

Figure~\ref{fig:event_time} visualises the pro-forma-to-live
transition in event time relative to each strategy's launch month.
The appendix is descriptive: it shows the shape of the decay
documented by the regressions in Section~\ref{sec:results} without
conditioning on pre-launch controls.  Panel~(a) shows the cumulative
equal-weighted mean excess return and Panel~(b) the rolling
six-month vol-adjusted return, both indexed to event month~0.
The magnitude of the discontinuity at launch is consistent with the
coefficient estimates in the main decay regressions.

\section{Bootstrap Block-Size Sensitivity}
\label{app:bootstrap_sensitivity}

The bootstrap targets a simplified univariate regression of decay on
the horizon-matched pro-forma vol-adjusted return (winsorized at
1\%/99\%, no fixed effects), which delivers an observed coefficient of
$\hat\beta_{\text{decay}} = -0.801$.  This differs in magnitude from
the FE-controlled baseline estimate of $-0.896$ in
Table~\ref{tab:bench_summary_decay}: the bootstrap sacrifices the
controls to make the strategy-level block resampling tractable, but
identifies the same mean-reversion-beyond-sampling-noise effect.
Table~\ref{tab:bootstrap_block_sensitivity} repeats the bootstrap null
exercise under four alternative block sizes
(10, 21, 42, and 63 trading days).
The observed excess negativity ($-0.06$ to $-0.10$) and the $p$-value
($\le 0.004$ throughout) are stable across all block choices.
Across the full range of block sizes tested, the observed decay
coefficient falls well outside the 90\% null confidence interval, and
the associated $p$-value never rises above $0.004$.  The decay result
is therefore not an artefact of a particular bootstrap block choice,
and the moving-block bootstrap inference reported in
Section~\ref{sec:results} is insensitive to this tuning parameter.

  \begingroup
  \singlespacing
  \scriptsize
  \setlength{\tabcolsep}{3pt}%
  \renewcommand{\arraystretch}{0.90}%
  \input{table_bootstrap_block_sensitivity}%
  \endgroup

\TableBarrier

\section{Placebo Launch-Timing Tests}
\label{app:placebo}

The decay documented in Section~\ref{sec:results} is consistent with
at least two distinct selection mechanisms.  Under a
bucket-regime channel, launches cluster at temporary peaks of
each strategy's bucket-factor environment; under a
within-strategy channel, each strategy is launched at its own
idiosyncratic high-water mark on a pre-launch performance signal.
The main text argues for the first channel.  This appendix rules out
the second by means of a per-strategy permutation test.

For each live strategy $i$ with launch month $t_i$, we define a
36-month pre-launch window $[t_i - 36,\, t_i - 1]$ and compute, at
$t_i$, the six-month trailing volatility-adjusted return as the
pre-launch signal $s_i$.  We then draw $B = 999$ placebo months
uniformly at random, with replacement, from the non-launch months
of the same window, and evaluate the pre-launch signal at each
placebo month.  This generates, for every
strategy, an empirical null distribution of the pre-launch signal
under the assumption that the launch date is unrelated to
within-strategy peaks.  A per-strategy one-sided $p$-value is obtained as
the rank of $s_i$ in its own placebo distribution.  The test is
run on the paper's baseline analysis sample, so the placebo sample
of 1{,}726 strategies coincides exactly with the decay-regression
sample of Section~\ref{sec:results}.

We aggregate the per-strategy evidence in two complementary ways.
First, a two-sided one-sample $t$-test on per-strategy lifts (the
real-launch signal minus the placebo mean) tests the null that mean
launch-date signals equal mean pre-launch signals.  Second, we
tabulate the fraction of strategies with per-strategy $p < 0.10$
and $p < 0.05$ and compare these to their nominal sizes; systematic
cherry-picking would produce right-skewed $p$-values and rejection
rates well above 10\% and 5\%.

Across $N = 1{,}726$ live strategies, the mean per-strategy lift is
$-0.016$ ($p < 0.001$): real launches arrive
slightly below the placebo mean on the pre-launch vol-adjusted
return.  At the per-strategy level, 9.8\% of strategies reject the
placebo null at $p < 0.10$ and 5.3\% at $p < 0.05$, essentially at
their nominal levels of 10\% and 5\%.

Launches are therefore not systematically timed to within-strategy
peaks; if anything, they are marginally below each strategy's own
pre-launch mean.  This rules out local peak-picking by individual
strategies as the driver of the decay documented in
Section~\ref{sec:results} and is consistent with the bucket-regime
channel formalised in Section~\ref{sec:regime_extremity}: launches
may still be timed to common bucket-factor peaks while remaining
idiosyncratically neutral within each strategy's own history.

\section{Wild Cluster Bootstrap P-Values}
\label{app:wild_bootstrap}

Table~\ref{tab:wild_bootstrap} reports wild cluster restricted (WCR)
bootstrap $p$-values for the key coefficients under
institution-level clustering (ten clusters).
Following \citet{Cameron2015}, we use the Webb 6-point distribution
with 999 replications.
The bootstrap generates data under $H_0\colon \beta_{\text{key}} = 0$
using restricted fitted values and restricted residuals, then
compares the distribution of bootstrap $t$-statistics to the observed
$t$-statistic (two-sided).
Both the six-month and twelve-month decay coefficients are highly
significant under the wild bootstrap ($p < 0.001$ at each horizon).
The live-return coefficient is also significant at both horizons
($p = 0.005$ at six months; $p = 0.011$ at twelve months).
All four key coefficients are robust to the conservatism inherent
in ten-cluster inference.

  \begingroup
  \singlespacing
  \scriptsize
  \setlength{\tabcolsep}{3pt}%
  \renewcommand{\arraystretch}{0.90}%
  \input{table_wild_bootstrap}  \endgroup

\section{Bloomberg Benchmark: Single- and Two-Factor Specifications}
\label{app:bloomberg_twofactor}

Table~\ref{tab:benchmark_live_alpha} reports the full Bloomberg
benchmark regressions under both the single-factor Jensen specification
(Equation~\ref{eq:jensen}) and the two-factor variant augmented with
the VIX term-structure convexity proxy
(Equation~\ref{eq:jensen_convex}).
Conclusions are identical across both specifications.
Both horizons are estimated on the intersection sample of $N = 1{,}694$
strategies with valid twelve-month live observations, so that the 6m
and 12m rows use an identical universe and differ only in the
horizon-matched outcome.

  \begingroup
  \singlespacing
  \tiny
  \setlength{\tabcolsep}{2.5pt}%
  \renewcommand{\arraystretch}{0.84}%
  \input{table8_benchmark_live_alpha}%
  \endgroup

\section{Live-Only LOO Benchmark}
\label{app:loo_liveonly}

The LOO bucket-average benchmark defined in Equation~\ref{eq:loo_def}
includes returns from all peer strategies on each date, whether they
are in their pro-forma or live period.
This is the natural choice when the benchmark is intended to capture the
realised bucket-factor regime, but a referee might reasonably ask
whether including backtest returns introduces look-ahead bias into the
peer average.
To address this concern, we re-estimate the LOO benchmark using only
strategies that have already launched (i.e., date~$\geq$~live\_date)
on each trading day, setting non-live returns to missing before computing
the daily bucket average.

The core decay result (Table~\ref{tab:bench_summary_decay}, LOO row) is
virtually unchanged: the coefficient on pro-forma vol-adjusted return in
the LOO-relative decay regression is $-0.358$ (SE~$= 0.019$,
$p < 0.001$) under the live-only benchmark, compared with $-0.372$
(SE~$= 0.019$, $p < 0.001$) in the baseline---both significant at
the 1\% level.
The LOO-relative live-performance coefficient
(Table~\ref{tab:bench_summary_levels}, LOO row) is similarly stable at
$0.032$ (SE~$= 0.013$, $p = 0.014$) versus the baseline
$0.034$ (SE~$= 0.013$, $p = 0.010$).

We further verify the regime-extremity channel test
(Table~\ref{tab:regime_extremity}, Specification~iii) under the most
conservative variant, which combines the live-only LOO benchmark with
a LOO bucket mean in the regime-extremity computation
(excluding strategy~$i$ from its own bucket average).
Under this specification, the regime-extremity coefficient
strengthens: $\hat\beta = 0.827$ (SE~$= 0.095$, $p < 0.001$) at
twelve months and $\hat\beta = 0.518$ (SE~$= 0.142$, $p < 0.001$)
at six months, compared with the baseline estimates of $0.716$ and
$0.469$ respectively.
The $R^2$ of the twelve-month specification rises from 0.293 to
0.319.
These results confirm that the baseline LOO benchmark, if anything,
yields conservative estimates of both the decay pattern and the
regime-timing channel.

\section{Raw vs.\ Benchmark-Adjusted Coefficients: Matched-Sample Test}
\label{app:raw_vs_adjusted}

Table~\ref{tab:raw_vs_adjusted_test} tests whether the attenuation from
raw to benchmark-adjusted coefficients is a substantive change in the
economic relationship or a mechanical consequence of using a noisier
dependent variable.
On the identical matched sample, the raw coefficient remains
economically large while the benchmark-adjusted coefficient collapses
toward zero; the Wald test confirms that the difference is significant
($z = 4.03$, $p < 0.001$).

  \begingroup
  \singlespacing
  \scriptsize
  \setlength{\tabcolsep}{3pt}%
  \renewcommand{\arraystretch}{0.90}%
  \input{table_raw_vs_adjusted_test}  \endgroup

\TableBarrier

\section{Benchmark-Relative Performance: Distributional Detail}
\label{app:bench_rel_dist}

Table~\ref{tab:bench_rel_dist} reports the full distribution of
benchmark-relative live performance (Panel~A) and benchmark-relative
decay (Panel~B) under the three benchmark conventions discussed in
Section~\ref{sec:loo_results}: no adjustment (raw vol-adjusted
return), the simple Bloomberg-relative return, and the simple
LOO-relative return.
The table underpins the numerical statements made at the start of
Section~\ref{sec:loo_results}.

\input{table_benchmark_relative_distribution}
\TableBarrier

\end{appendices}

\end{document}

%% file: table1_summary_main.tex
\begin{table}
\caption{Sample Summary. Panel~A: Composition by asset class, objective group, and bucket. Panel~B: Distribution of track-record lengths (years). The full institution-level and launch-year breakdown is reported in Appendix~\ref{app:sample_detail}.}
\label{tab:sample_summary}
Panel A: Composition\\[0.3em]
\begin{tabular}{l r r @{\hspace{1em}} l r r @{\hspace{1em}} l r r}
\toprule
Group & N & Share & Group & N & Share & Group & N & Share \\
\midrule
Total strategies & 1726 & 100\% & {By objective group} &  &  & {By bucket} &  &  \\
{By asset class} &  &  & \quad Return-Seeking & 457 & 26\% & \quad Carry & 808 & 47\% \\
\quad Commodities & 366 & 21\% & \quad Carry/Short-Convexity & 808 & 47\% & \quad Momentum & 215 & 12\% \\
\quad Credit & 132 & 8\% & \quad Hedging/Defensive & 300 & 17\% & \quad Hedging & 300 & 17\% \\
\quad Equities & 670 & 39\% & \quad Multi-Premia/Div & 161 & 9\% & \quad Multi Premia & 161 & 9\% \\
\quad FX & 197 & 11\% &  &  &  & \quad Factor & 154 & 9\% \\
\quad Multi-Asset & 118 & 7\% &  &  &  & \quad Value & 59 & 3\% \\
\quad Rates & 243 & 14\% &  &  &  & \quad Liquidity & 29 & 2\% \\
\bottomrule
\end{tabular}
\vspace{0.8em}

Panel B: Track-record lengths\\[0.3em]
\begin{tabular}{l r r r r r r}
\toprule
Variable & N & Mean & Std & P25 & Median & P75 \\
\midrule
Pro-forma period (years) & 1726 & 13.7 & 6.4 & 10.4 & 14.3 & 17.3 \\
Live period (years) & 1726 & 5.6 & 3.2 & 3.2 & 5.2 & 7.6 \\
\bottomrule
\end{tabular}
\end{table}

%% file: table2_pf_vs_live_a.tex
\begin{table}[htbp]
\centering
\caption{Pro-Forma vs Live Performance Summary. All figures are in per cent p.a.\ (e.g.\ 3.6 means 3.6\% p.a.). Each cell reports the mean vol-adjusted return or decay (live minus pro-forma) over 6-month and 12-month post-launch windows. Returns are rescaled to a common 10\% annualised volatility target. Sample sizes are horizon-specific: the $N$ column under the 6-month horizon counts strategies with at least six months of live data, and the $N$ column under the 12-month horizon counts strategies with at least twelve months. \textsuperscript{*}\,$p<0.10$, \textsuperscript{**}\,$p<0.05$, \textsuperscript{***}\,$p<0.01$ for the test $H_0$: decay $= 0$.}
\label{tab:pf_vs_live_a}
\small
\begin{tabular}{l r rr r@{}l r rr r@{}l}
\toprule
 & \multicolumn{5}{c}{6-month horizon} & \multicolumn{5}{c}{12-month horizon} \\
\cmidrule(lr){2-6} \cmidrule(lr){7-11}
Panel & $N$ & Pre (\%) & Live (\%) & \multicolumn{2}{c}{Decay (\%)} & $N$ & Pre (\%) & Live (\%) & \multicolumn{2}{c}{Decay (\%)} \\
\midrule
Full sample          & 1,726 & 3.6 & 1.5 & $-$2.1 & *** & 1,694 & 4.1 & 1.0 & $-$3.1 & *** \\[0.3em]
Commodities          &   366 & 7.1 & 3.1 & $-$4.0 & *** &   358 & 6.5 & 2.1 & $-$4.6 & *** \\[0.3em]
Credit               &   132 & $-$0.3 & $-$0.4 & $-$0.1 &  &   132 & 2.8 & $-$0.5 & $-$3.4 & ** \\[0.3em]
Equities             &   670 & 3.6 & 1.9 & $-$1.7 & ** &   656 & 4.1 & 1.1 & $-$3.0 & *** \\[0.3em]
FX                   &   197 & 3.4 & $-$1.1 & $-$4.6 & *** &   195 & 3.4 & $-$0.1 & $-$3.5 & *** \\[0.3em]
Multi-Asset          &   118 & 2.2 & 1.4 & $-$0.8 &  &   115 & 3.5 & 0.5 & $-$3.0 & ** \\[0.3em]
Rates                &   243 & 1.1 & 1.3 & 0.3 &  &   238 & 1.9 & 1.1 & $-$0.7 &  \\
\bottomrule
\end{tabular}
\end{table}

%% file: table_benchmark_summary_levels.tex
\begin{table}[htbp]
\caption{Pro-Forma Predictability of Live Performance Across Benchmarks. Each row reports the OLS coefficient on the key pro-forma predictor under a different benchmark specification. Row~1 uses raw live vol-adjusted return as the dependent variable; Row~2 substitutes the OLS-estimated Jensen $\alpha$ against the Bloomberg total-return index; Row~3 substitutes the LOO-relative return. All regressions include bucket $\times$ launch-year fixed effects; standard errors clustered by strategy. \textsuperscript{***}\,p<0.01, \textsuperscript{**}\,p<0.05, \textsuperscript{*}\,p<0.10.}
\label{tab:bench_summary_levels}
\centering
\begin{tabular}{llcccrc}
\toprule
Benchmark & Dependent variable & $\hat{\beta}$ & SE & $t$ & $N$ & $R^2$ \\
\midrule
None (raw) & Live return (12m, vol-adj.) & 0.137*** & (0.025) & 5.48 & 1694 & 0.148 \\
Bloomberg index & Jensen alpha (12m) & 0.025** & (0.012) & 2.05 & 1694 & 0.032 \\
LOO relative return & LOO relative return (12m) & 0.034*** & (0.013) & 2.58 & 1694 & 0.054 \\
\bottomrule
\end{tabular}
\end{table}

%% file: table_benchmark_summary_decay.tex
\begin{table}[htbp]
\caption{Pro-Forma Predictability of Performance Decay Across Benchmarks. Each row reports the OLS coefficient on the key pro-forma predictor under a different benchmark specification. Row~1 uses raw vol-adjusted return decay as the dependent variable; Row~2 uses Bloomberg-relative decay (live minus pro-forma benchmark-relative return); Row~3 uses LOO-relative decay. All regressions include bucket $\times$ launch-year fixed effects; standard errors clustered by strategy. \textsuperscript{***}\,p<0.01, \textsuperscript{**}\,p<0.05, \textsuperscript{*}\,p<0.10.}
\label{tab:bench_summary_decay}
\centering
\begin{tabular}{llcccrc}
\toprule
Benchmark & Dependent variable & $\hat{\beta}$ & SE & $t$ & $N$ & $R^2$ \\
\midrule
None (raw) & Return decay (12m, vol-adj.) & $-$0.896*** & (0.025) & $-$35.81 & 1694 & 0.494 \\
Bloomberg relative & Bloomberg-relative decay (12m) & $-$0.377*** & (0.031) & $-$12.01 & 1694 & 0.247 \\
LOO bucket-average & LOO-relative decay (12m) & $-$0.372*** & (0.019) & $-$19.46 & 1694 & 0.267 \\
\bottomrule
\end{tabular}
\end{table}

%% file: table_regime_extremity_paper.tex
\begin{table}[htbp]
\caption{Regime Extremity and Performance Decay.
         Each column group tests whether the intensity of the bucket-factor
         regime at launch predicts subsequent performance decay.
         Specifications~(i)--(ii) use \textbf{raw} vol-adjusted return decay
         ($\Delta r^{\text{adj}}_{i,h}$, Equation~\ref{eq:decay_def})
         as the dependent variable: (i)~adds regime extremity as an
         additive control; (ii)~further adds the interaction of pro-forma
         return with regime extremity (\textbf{Pre-ret $\times$ regime}),
         testing whether hotter regimes amplify the pro-forma signal.
         Specification~(iii) replaces the dependent variable with
         \textbf{LOO-relative} decay ($\Delta r^{\text{LOO}}_{i,h}$,
         Equation~\ref{eq:loo_rel_decay}), which strips out the common
         bucket factor and isolates strategy-specific deterioration.
         \textbf{Regime extremity} is the annualised LOO bucket-average
         return in the 12~months before launch minus the within-bucket
         full-sample mean (Equation~\ref{eq:regime_extremity}).
         All specifications include bucket $\times$ launch-year fixed effects
         and use the horizon-matched pro-forma vol-adjusted return
         ($r^{\text{adj,pre}}_{i,h}$) as the key predictor.
         All specifications also control for earlier-window vol-adjusted
         return, pro-forma volatility, and strategy age at launch
         (Equation~\ref{eq:controls}); these coefficients are omitted from
         display for compactness and are available on request.
         Standard errors (in parentheses) are clustered by strategy.
         \textsuperscript{***}\,$p<0.01$; \textsuperscript{**}\,$p<0.05$; \textsuperscript{*}\,$p<0.10$.}
\label{tab:regime_extremity}
\centering
\footnotesize
\begin{tabular}{l *{6}{c}}
\toprule
 & \multicolumn{2}{c}{(i) Additive} & \multicolumn{2}{c}{(ii) + Interaction} & \multicolumn{2}{c}{(iii) LOO-relative} \\
\cmidrule(lr){2-3} \cmidrule(lr){4-5} \cmidrule(lr){6-7}
 & 12m & 6m & 12m & 6m & 12m & 6m \\
\midrule
Regime extremity$_{12\text{m}}$ & $-0.087$ & $0.206$ & $-0.003$ & $0.083$ & $0.716^{***}$ & $0.469^{***}$ \\
 & $(0.132)$ & $(0.182)$ & $(0.152)$ & $(0.199)$ & $(0.124)$ & $(0.176)$ \\[4pt]
Pre-ret $\times$ regime &  &  & $-1.637$ & $4.580^{***}$ &  &  \\
 &  &  & $(1.296)$ & $(1.446)$ &  &  \\[4pt]
Pro-forma vol-adj return$_{h}$ & $-0.891^{***}$ & $-0.833^{***}$ & $-0.894^{***}$ & $-0.827^{***}$ & $-0.409^{***}$ & $-0.342^{***}$ \\
 & $(0.026)$ & $(0.029)$ & $(0.026)$ & $(0.028)$ & $(0.019)$ & $(0.020)$ \\[4pt]
\midrule
Bucket $\times$ Launch-year FE & \cmark & \cmark & \cmark & \cmark & \cmark & \cmark \\
Clustered SE & \cmark & \cmark & \cmark & \cmark & \cmark & \cmark \\
$N$ & 1{,}694 & 1{,}726 & 1{,}694 & 1{,}726 & 1{,}694 & 1{,}726 \\
$R^2$ & 0.494 & 0.423 & 0.495 & 0.427 & 0.293 & 0.228 \\
\bottomrule
\end{tabular}
\end{table}

%% file: table_launch_density_paper.tex
\begin{table}[htbp]
\caption{Launch-Cohort Density and Performance Decay.
         Each column group tests a different aspect of the crowding channel.
         Specifications~(A) and~(B) use \textbf{raw} vol-adjusted return
         decay ($\Delta r^{\text{adj}}_{i,h}$, Equation~\ref{eq:decay_def})
         as the dependent variable: (A)~asks whether launching into a more
         crowded bucket-year directly predicts worse raw decay;
         (B)~adds the interaction of pro-forma return with log density
         (\textbf{Pre-ret $\times$ density}), testing whether crowding
         \textbf{amplifies} the pro-forma signal rather than acting as a
         level shift.
         Specification~(C) replaces the dependent variable with
         \textbf{LOO-relative} decay ($\Delta r^{\text{LOO}}_{i,h}$,
         Equation~\ref{eq:loo_rel_decay}), which strips out the common
         bucket factor and isolates strategy-specific deterioration
         (directly analogous to Specification~(iii) of the regime-extremity
         table).
         \textbf{Log launch density} is $\ln(1 + n_{b,y})$ where $n_{b,y}$
         is the number of strategies launched in bucket~$b$ during year~$y$
         (Equation~\ref{eq:launch_density}).
         All specifications include bucket fixed effects only.
         All specifications use the horizon-matched pro-forma vol-adjusted
         return ($r^{\text{adj,pre}}_{i,h}$) as the key predictor.
         Standard errors (in parentheses) are clustered by strategy.
         \textsuperscript{***}\,$p<0.01$; \textsuperscript{**}\,$p<0.05$; \textsuperscript{*}\,$p<0.10$.}
\label{tab:launch_density}
\centering
\footnotesize
\begin{tabular}{l *{6}{c}}
\toprule
 & \multicolumn{2}{c}{(A) Density} & \multicolumn{2}{c}{(B) + Interaction} & \multicolumn{2}{c}{(C) LOO-relative} \\
\cmidrule(lr){2-3} \cmidrule(lr){4-5} \cmidrule(lr){6-7}
 & 12m & 6m & 12m & 6m & 12m & 6m \\
\midrule
Log launch density & $-0.000$ & $0.001$ & $-0.001$ & $-0.004$ & $-0.004$ & $0.006$ \\
 & $(0.005)$ & $(0.006)$ & $(0.005)$ & $(0.007)$ & $(0.003)$ & $(0.004)$ \\[4pt]
Pre-ret $\times$ density &  &  & $0.023$ & $0.109^{***}$ &  &  \\
 &  &  & $(0.026)$ & $(0.026)$ &  &  \\[4pt]
Pro-forma vol-adj return$_{h}$ & $-0.897^{***}$ & $-0.824^{***}$ & $-0.983^{***}$ & $-1.211^{***}$ & $-0.372^{***}$ & $-0.329^{***}$ \\
 & $(0.025)$ & $(0.028)$ & $(0.095)$ & $(0.093)$ & $(0.019)$ & $(0.019)$ \\[4pt]
\midrule
Bucket FE & \cmark & \cmark & \cmark & \cmark & \cmark & \cmark \\
Clustered SE & \cmark & \cmark & \cmark & \cmark & \cmark & \cmark \\
$N$ & 1{,}694 & 1{,}726 & 1{,}694 & 1{,}726 & 1{,}694 & 1{,}726 \\
$R^2$ & 0.482 & 0.410 & 0.482 & 0.416 & 0.249 & 0.201 \\
\bottomrule
\end{tabular}
\end{table}

%% file: table_r3_failure_oos_compact.tex
\begin{table}
\caption{Out-of-Sample Failure Prediction: Summary (Time-Split Validation). Selected out-of-sample failure outcomes by objective group, estimated from the classifier defined in Equation~\ref{eq:failure_classifier}. ``Base Rate'': unconditional failure frequency in the test cohort. ``AUC'': area under the ROC curve on the held-out later-launch cohort, interpretable as the probability that a randomly chosen failing strategy receives a higher predicted risk score than a randomly chosen non-failing one (0.5 = random classifier, 1.0 = perfect separation). ``Top-Decile Capture'': recall among strategies in the top 10\% of predicted risk -- the share of realised failures falling in the highest-risk decile (random baseline $= 10\%$). ``Top-Decile Rate'': precision among strategies in the top 10\% of predicted risk -- the realised failure rate within the highest-risk decile. ``Lift'': ratio of Top-Decile Rate to Base Rate.}
\label{tab:classifier_oos}
\resizebox{\textwidth}{!}{
\begin{tabular}{ll r @{\hspace{1.8em}} r r r r}
\toprule
 &  & & \multicolumn{4}{c}{Model Performance (OOS)} \\
\cmidrule(l){4-7}
Group & Outcome & Base Rate & AUC & Top-Decile Capture & Top-Decile Rate & Lift \\
\midrule
Return-Seeking        & Negative Return    & 48.6\% & 0.477 & 7.1\%  & 35.3\% & 0.7x \\
Return-Seeking        & Performance Decay  & 25.6\% & 0.813 & 29.2\% & 76.5\% & 3.0x \\
Return-Seeking        & Multiple Failures  & 38.2\% & 0.776 & 19.5\% & 76.5\% & 2.0x \\
Carry/Short-Convexity & Negative Return    & 40.5\% & 0.551 & 12.5\% & 50.8\% & 1.3x \\
Carry/Short-Convexity & Multiple Failures  & 37.0\% & 0.608 & 11.9\% & 44.4\% & 1.2x \\
Hedging/Defensive     & Multiple Failures  & 40.9\% & 0.567 & 13.4\% & 55.6\% & 1.4x \\
Multi-Premia/Diversifying & Multiple Failures & 37.8\% & 0.639 & 14.3\% & 54.5\% & 1.4x \\
\bottomrule
\end{tabular}
}
\end{table}

%% file: table_bootstrap_block_sensitivity.tex
\begin{table}[htbp]
\caption{Sensitivity of Bootstrap Null to Block Size. The observed decay coefficient $\hat{\beta}_{\text{decay}} = -0.801$ is compared against the bootstrap null distribution under four alternative block sizes (10, 21, 42, 63 trading days). Null mean and 90\% CI are from 500 bootstrap replications per block size. Excess negativity = observed minus null mean. $p$-value is the fraction of null draws $\leq$ the observed coefficient. The excess negativity and significance are stable across block sizes.}
\label{tab:bootstrap_block_sensitivity}
\begin{tabular}{lrrrrrr}
\toprule
Block size (days) & N reps & Null mean & Null 90\% CI & Observed $\hat{{\beta}}$ & Excess negativity & $p$-value \\
\midrule
10 & 500 & -0.740 & [-0.779, -0.702] & -0.801 & -0.060 & 0.0040 \\
21 & 500 & -0.725 & [-0.765, -0.684] & -0.801 & -0.075 & $<0.002$ \\
42 & 500 & -0.711 & [-0.753, -0.671] & -0.801 & -0.089 & $<0.002$ \\
63 & 500 & -0.702 & [-0.743, -0.665] & -0.801 & -0.099 & $<0.002$ \\
\bottomrule
\end{tabular}
\end{table}

%% file: table_wild_bootstrap.tex
\begin{table}
\caption{Wild Cluster Bootstrap P-Values (Institution-Level Clustering). Webb 6-point distribution with 999 replications. Null hypothesis: coefficient on the key predictor equals zero. Restricted-residual bootstrap under H$_0$. With only ten institution clusters, conventional cluster-robust standard errors may be unreliable \citep{Cameron2015}; these wild bootstrap $p$-values provide a more reliable finite-cluster inference.}
\label{tab:wild_bootstrap}
\resizebox{\textwidth}{!}{
\begin{tabular}{llrrrrrl}
\toprule
Outcome & Key predictor & Observed coef & Observed t & N clusters & N obs & Boot replications & Wild bootstrap p \\
\midrule
Vol-adj return decay (12m) & Pro-forma vol-adj return (12m) & -0.896 & -38.708 & 10 & 1694 & 999 & $<0.001$ \\
Vol-adj return decay (6m) & Pro-forma vol-adj return (6m) & -0.828 & -33.400 & 10 & 1726 & 999 & $<0.001$ \\
Live vol-adj return (12m) & Pro-forma vol-adj return (12m) & 0.137 & 5.932 & 10 & 1694 & 999 & 0.011 \\
Live vol-adj return (6m) & Pro-forma vol-adj return (6m) & 0.210 & 8.444 & 10 & 1726 & 999 & 0.005 \\
\bottomrule
\end{tabular}
}
\end{table}

%% file: table8_benchmark_live_alpha.tex
\begin{table}
\caption{Bloomberg Benchmark: Single- vs.\ Two-Factor Jensen $\alpha$.
Each row reports the coefficient on the key pro-forma predictor
(horizon-matched vol-adjusted return) when the dependent variable is
Jensen $\alpha$ estimated under the single-factor specification
(Equation~\ref{eq:jensen}) or the two-factor specification augmented
with a VIX term-structure convexity proxy
(Equation~\ref{eq:jensen_convex}).
Both horizons are estimated on the intersection sample of
$N = 1{,}694$ strategies with valid twelve-month live observations,
so that the 6m and 12m rows share an identical universe.
All regressions include bucket $\times$ launch-year fixed effects;
standard errors clustered by strategy.
\textsuperscript{***}\,p<0.01, \textsuperscript{**}\,p<0.05, \textsuperscript{*}\,p<0.10.}
\label{tab:benchmark_live_alpha}
\centering
\begin{tabular}{llcccrc}
\toprule
Specification & Outcome & $\hat{\beta}$ & SE & $t$ & $N$ & $R^2$ \\
\midrule
\multicolumn{7}{l}{12-month horizon} \\[2pt]
(i) Pre-launch controls          & Jensen $\alpha$ (1F)   & 0.025**  & (0.012) &    2.05 & 1694 & 0.032 \\
(i) Pre-launch controls          & Jensen $\alpha$ (2F)   & 0.027**  & (0.012) &    2.21 & 1694 & 0.039 \\
(ii) + Idio.\ pro-forma signal   & Jensen $\alpha$ (1F)   & 0.010    & (0.012) &    0.80 & 1694 & 0.037 \\
(ii) + Idio.\ pro-forma signal   & Jensen $\alpha$ (2F)   & 0.012    & (0.012) &    1.01 & 1694 & 0.044 \\
(iii) + Benchmark run-up         & Jensen $\alpha$ (1F)   & 0.025**  & (0.012) &    2.07 & 1694 & 0.034 \\
(iii) + Benchmark run-up         & Jensen $\alpha$ (2F)   & 0.027**  & (0.012) &    2.22 & 1694 & 0.040 \\
(iv) Full                        & Jensen $\alpha$ (1F)   & $-$0.002 & (0.017) & $-$0.11 & 1694 & 0.042 \\
(iv) Full                        & Jensen $\alpha$ (2F)   & $-$0.003 & (0.018) & $-$0.18 & 1694 & 0.052 \\
\addlinespace
\multicolumn{7}{l}{6-month horizon} \\[2pt]
(i) Pre-launch controls          & Jensen $\alpha$ (1F)   & 0.025*   & (0.013) &    1.89 & 1694 & 0.046 \\
(i) Pre-launch controls          & Jensen $\alpha$ (2F)   & 0.027**  & (0.013) &    2.06 & 1694 & 0.049 \\
(ii) + Idio.\ pro-forma signal   & Jensen $\alpha$ (1F)   & $-$0.001 & (0.012) & $-$0.04 & 1694 & 0.058 \\
(ii) + Idio.\ pro-forma signal   & Jensen $\alpha$ (2F)   & 0.002    & (0.012) &    0.18 & 1694 & 0.061 \\
(iii) + Benchmark run-up         & Jensen $\alpha$ (1F)   & 0.024*   & (0.013) &    1.81 & 1694 & 0.054 \\
(iii) + Benchmark run-up         & Jensen $\alpha$ (2F)   & 0.026**  & (0.013) &    1.99 & 1694 & 0.056 \\
(iv) Full                        & Jensen $\alpha$ (1F)   & $-$0.006 & (0.016) & $-$0.39 & 1694 & 0.063 \\
(iv) Full                        & Jensen $\alpha$ (2F)   & $-$0.007 & (0.016) & $-$0.43 & 1694 & 0.067 \\
\bottomrule
\end{tabular}
\end{table}

%% file: table_raw_vs_adjusted_test.tex
\begin{table}
\caption{Formal Comparison: Raw vs.\ Benchmark-Controlled vs.\ Benchmark-Adjusted Predictability. All three regressions estimated on the identical benchmark-matched subsample. ``Raw'': live vol-adjusted return as outcome, no benchmark control. ``+Bench ctrl'': live vol-adjusted return as outcome with Bloomberg benchmark return added as RHS control (single-step). ``Jensen $\alpha$'': benchmark-adjusted alpha as outcome (two-step). Wald $z$: test of coefficient equality (conservative, assuming independence). Bucket $\times$ Launch-year FE throughout.}
\label{tab:raw_vs_adjusted_test}
\resizebox{\textwidth}{!}{
\begin{tabular}{lrrrrrrrrrrrlrl}
\toprule
Horizon & N & Coef (raw) & SE (raw) & R2 (raw) & Coef (+bench ctrl) & SE (ctrl) & R2 (ctrl) & Coef (Jensen alpha) & SE (alpha) & R2 (alpha) & Wald z (raw vs ctrl) & p (raw vs ctrl) & Wald z (raw vs alpha) & p (raw vs alpha) \\
\midrule
6m & 1726 & 0.210 & 0.028 & 0.151 & 0.214 & 0.028 & 0.155 & 0.025 & 0.013 & 0.042 & -0.103 & 0.918 & 5.963 & $<0.001$ \\
12m & 1694 & 0.137 & 0.025 & 0.148 & 0.142 & 0.025 & 0.155 & 0.025 & 0.012 & 0.032 & -0.146 & 0.884 & 4.030 & $<0.001$ \\
\bottomrule
\end{tabular}
}
\end{table}

%% file: table_benchmark_relative_distribution.tex
\begin{table}[htbp]
\caption{Distribution of Benchmark-Relative Live Performance and Decay. All values are in percentage points p.a. Panel~A reports the distribution of the benchmark-relative live return (the live vol-adjusted return after subtracting the benchmark return over the same window) at the six- and twelve-month horizons under three benchmark conventions: none (raw vol-adjusted live return), Bloomberg (live return minus assigned Bloomberg total-return index), and LOO (live return minus leave-one-out bucket-peer average). Panel~B reports the distribution of benchmark-relative decay (live minus pro-forma benchmark-relative return). The column ``\% neg'' is the fraction of strategies with a negative value (for levels, underperformance relative to the benchmark; for decay, deterioration relative to the pro-forma window).}
\label{tab:bench_rel_dist}
\centering
\small
\begin{tabular}{l l r r r r r r r}
\toprule
Benchmark & Horizon & $N$ & Mean & Std & P25 & Median & P75 & \% neg \\
\midrule
\multicolumn{9}{l}{\textit{Panel A: Benchmark-relative live return (percentage points p.a.)}} \\
\midrule
None (raw)        & 6m  & 1{,}726 & $+$1.5 & 16.7 & $-$9.5 & $+$0.9 & $+$12.2 & 47.0\% \\
None (raw)        & 12m & 1{,}694 & $+$1.0 & 11.7 & $-$6.8 & $+$1.1 & $+$8.3 & 46.6\% \\
Bloomberg& 6m  & 1{,}726 & $-$3.1 & 21.0 & $-$16.3 & $-$2.8 & $+$10.6 & 56.3\% \\
Bloomberg& 12m & 1{,}694 & $-$3.6 & 14.7 & $-$12.9 & $-$3.0 & $+$5.7 & 59.2\% \\
LOO& 6m  & 1{,}726 & $-$0.4 & 8.5 & $-$4.0 & $-$0.7 & $+$2.8 & 55.1\% \\
LOO& 12m & 1{,}694 & $-$0.8 & 6.2 & $-$3.6 & $-$0.8 & $+$1.7 & 59.1\% \\
\midrule
\multicolumn{9}{l}{\textit{Panel B: Benchmark-relative decay (live minus pro-forma, percentage points p.a.)}} \\
\midrule
None (raw)        & 6m  & 1{,}726 & $-$2.1 & 20.3 & $-$14.4 & $-$3.2 & $+$10.5 & 57.1\% \\
None (raw)        & 12m & 1{,}694 & $-$3.1 & 15.2 & $-$13.2 & $-$3.5 & $+$6.4 & 58.8\% \\
Bloomberg& 6m  & 1{,}726 & $+$0.3 & 20.3 & $-$9.8 & $+$1.5 & $+$11.1 & 46.3\% \\
Bloomberg& 12m & 1{,}694 & $-$1.4 & 16.1 & $-$9.1 & $-$0.6 & $+$7.7 & 52.8\% \\
LOO& 6m  & 1{,}726 & $-$0.6 & 11.7 & $-$5.0 & $-$0.7 & $+$3.0 & 54.7\% \\
LOO& 12m & 1{,}694 & $-$1.3 & 9.3 & $-$4.7 & $-$0.7 & $+$2.3 & 55.8\% \\
\bottomrule
\end{tabular}
\end{table}